\documentclass[journal]{IEEEtran}
\usepackage{cite}
\usepackage{url}
\usepackage{empheq}
\usepackage{float}
\usepackage{dsfont}
\usepackage[hidelinks]{hyperref}

\usepackage{algorithm}
\usepackage{algorithmic}

\usepackage{amsmath,amssymb,amsfonts}
\allowdisplaybreaks[4]
\usepackage{graphicx}
\usepackage{textcomp}
\usepackage{amsmath}
\usepackage{enumerate}
\usepackage{epstopdf}
\usepackage{array}
\usepackage{booktabs}
\usepackage{subfigure}
\usepackage{multirow}
\usepackage{soul, color}
\usepackage[usenames,dvipsnames]{xcolor}
\usepackage[latin1]{inputenc}

\newtheorem{theorem}{Theorem}
\newtheorem{lemma}{Lemma}

\newtheorem{proposition}{Proposition}

\newtheorem{definition}{Definition}

\soulregister\cite7 
\soulregister\citep7 
\soulregister\citet7 
\soulregister\ref7
\soulregister\pageref7

\begin{document}

\title{Distributed Coordination of Charging Stations with Shared Energy Storage in a Distribution Network}

\author{
Dongxiang~Yan and
Yue~Chen,~\IEEEmembership{Member,~IEEE}

\thanks{This work was supported by the Chinese University of Hong Kong (CUHK) Direct Grant for Research No. 4055169. (Corresponding to Y. Chen)}
\thanks{D. Yan and Y. Chen are with the Department of Mechanical and Automation Engineering, the Chinese University of Hong Kong, HKSAR, China (e-mail: dongxiangyan@cuhk.edu.hk, yuechen@mae.cuhk.edu.hk).
}}
\markboth{Journal of \LaTeX\ Class Files,~Vol.~XX, No.~X, Feb.~2019}%
{Shell \MakeLowercase{\textit{et al.}}: Bare Demo of IEEEtran.cls for IEEE Journals}

\maketitle

\begin{abstract}
Electric vehicle (EV) charging stations have experienced rapid growth, whose impacts on the power grid have become non-negligible. Though charging stations can install energy storage to reduce their impacts on the grid, the conventional ``one charging station, one energy storage'' method may be uneconomical due to the high upfront cost of energy storage. Shared energy storage can be a potential solution. However, effective management of charging stations with shared energy storage in a distribution network is challenging due to the complex coupling, competing interests, and information asymmetry between different agents. To address the aforementioned challenges, this paper first proposes an equilibrium model to characterize the interaction among charging stations, shared energy storage, and the distribution network. We prove that the equilibrium coincides with the centralized optimization result with trading prices equaling the value of dual variables at optimum. Then, to achieve the efficient equilibrium, a distributed coordination mechanism with a prediction and a correction step is developed to guide the behaviors of different agents with proof of convergence. Numerical experiments and comprehensive performance comparisons are conducted to validate the theoretical results and show the advantages of the proposed mechanism.
\end{abstract}


\IEEEpeerreviewmaketitle

\section*{Nomenclature}
\addcontentsline{toc}{section}{Nomenclature}
\subsection{Acronyms}
\begin{IEEEdescription}[\IEEEusemathlabelsep\IEEEsetlabelwidth{${\underline P _{mn}}$,${\overline P _{mn}}$}]
\item[ADMM]{Alternating direction method of multipliers.}
\item[CASAP]{Charging as soon as possible.}
\item[CS]{Charging station.}
\item[CSO]{Charging station operator.}
\item[DSO]{Distribution system operator.}
\item[EV]{Electric vehicle.}
\item[PV]{Photovoltaic.}
\item[SE]{System equilibrium.}
\item[SES]{Shared energy storage.}
\item[SESO]{Shared energy storage operator.}
\end{IEEEdescription}

\subsection{Symbols}
\begin{IEEEdescription}[\IEEEusemathlabelsep\IEEEsetlabelwidth{$E_{v,i}^{min},E_{v,i}^{max}$}]
\item[$\mathcal{T},t$]{Set of time slots and index.}
\item[$\mathcal{N},n/j$]{Set of buses and index.}
\item[$\mathcal{E}$]{Set of lines.}
\item[$\Delta t$]{Time interval.}
\item[$\mathcal{I},i/j$]{Set of charging stations and index.}
\item[$\mathcal{V}_i,v$]{Set of EVs in charging station $i$ and index.}
\item[$\mathcal{B},b$]{Set of shared energy storage and index.}
\item[$\mathcal{I}_b$]{Set of CSs connected to SES $b$.}
\item[$\mathcal{D}_i$]{Set of constraints of charging demand for CS $i$.}
\item[$\mathcal{P}$]{Set of constraints of distribution network.}
\item[$\mathcal{F}_b$]{Set of constraints of SES $b$.}

\item[$t_{v,i}^a,t_{v,i}^d$]{EV $v$'s arrival time and departure time.}
\item[$E_{v,i}^{ini},E_{v,i}^{req}$]{EV $v$'s initial energy when it arrives and required energy when it leaves in CS $i$.}
\item[$p_{v,i}^{max}$]{EV $v$'s maximum charging power in CS $i$.}
\item[$E_{v,i}^{cha}$]{EV $v$'s Charged energy from $E_{v,i}^{ini}$ to $E_{v,i}^{req}$.}
\item[$t_{v,i}^{min}$]{Minimum time needed for EV $v$ to be charged to its required energy level $E_{v,i}^{req}$}
\item[$C_{cs,i}$]{Combined EV inconvenience charging cost and EV battery depreciation cost in CS $i$.}
\item[$C_{cso,i}$]{Overall operation cost of CSO $i$.}
\item[$c_{v,i}$]{Inconvenience cost coefficient of EV $v$.}
\item[$\bar{p}_{v,i}$]{EV $v$'s desired charging power in CS $i$.}
\item[$p_{v,i,t}$]{EV $v$'s scheduled power at time $t$ in CS $i$.}
\item[$c_{v,i}^{dep}$]{Depreciation cost coefficient of EV $v$.}
\item[$\eta_{c,v}/\eta_{d,v}$]{Charging/discharging efficiency of EV $v$.}
\item[$p_{v,i,t}^{c}/p_{v,i,t}^{d}$]{EV $v$'s charging/discharging power at time $t$ in CS $i$.}
\item[$E_{v,i,t}$]{EV $v$'s energy at time $t$ in CS $i$.}
\item[$E_{v}^{min},E_{v}^{max}$]{Minimum/maximum value of $E_{v,i,t}$.}
\item[$p_{d,i,t}$]{Total charging power in CS $i$ at time $t$.}
\item[$p_{pv,i,t}$]{PV power generation in CS $i$ at time $t$.}
\item[$p_{g,i,t},\lambda_{g,i,t}$]{Selling power from CS $i$ to grid at time $t$, and the trading price.}
\item[$p_{b,i,t},\lambda_{b,i,t}$]{Selling power from CS $i$ to SES at time $t$, and the trading price.}
\item[$C_{bat,b}$]{Degradation cost of SES $b$.}
\item[$C_{seso,b}$]{Overall operation cost of SESO $b$.}
\item[$c_{b}$]{SES degradation coefficient.}
\item[$p^d_{b,i,d},p^c_{b,i,t}$]{Discharging/charging power from SES for CS $i$.}
\item[$p^d_{b,g,d},p^c_{b,g,t}$]{Discharging/charging power from SES for grid.}
\item[$p^{d,max}_{b},p^{c,max}_{b}$]{Maximum discharging and charging power for SES $b$.}
\item[$\eta_c,\eta_d$]{Charging/discharging efficiency of shared energy storage.}
\item[$E_{b,t}$]{SES $b$'s energy at time slot $t$.}
\item[$E_{b,cap}$]{SES $b$'s full energy capacity.}
\item[$E^{min}_{b},E^{max}_{b}$]{Minimum/maximum value of $E_{b,i,t}$.}
\item[$\lambda_{b,t}$]{Electricity buying price by grid from SES $b$ at time $t$.}
\item[$p_{j},q_{j}$]{Active and reactive power at bus $j$}
\item[$P_{nj},Q_{nj}$]{Active and reactive power flow in line $(n,j)$}
\item[$r_{nj},x_{nj}$]{Resistance and reactance in line $(n,j)$.}
\item[$v_{j}$]{Squared voltage magnitude at bus $j$}
\item[$\ell_{nj}$]{Squared current magnitude in line $(n,j)$.}
\item[$p_{g,b,t}$]{Purchased power by grid from SES $b$.}
\item[$\lambda^b_{g,t},\lambda^s_{g,t}$]{Electricity buying/selling price from/to the utility grid for bus 1.}
\item[$p_{1,t}^{b},p_{1,t}^{s}$]{Energy bought/sold from/to utility grid by bus 1 at time $t$.}
\item[$C_{ds}$]{Utility grid cost for DSO.}
\item[$C_{dso}$]{Overall operation cost of DSO.}

\item[$\lambda_{i,t},\mu_{b,t}$]{Dual variables.}
\item[$\boldsymbol{\lambda},\boldsymbol{\mu}$]{Matrix of dual variables $\lambda_{i,t},\mu_{b,t}$.}
\item[$k$]{Iteration.}
\item[$\beta,\delta$]{Penalty parameter, accuracy tolerance.}
\item[$\alpha,\tau$]{Parameters.}
\end{IEEEdescription}

\section{Introduction}
\IEEEPARstart {D}{riven} by the proliferation of electric vehicles (EVs), charging stations are expanding quickly \cite{wang2018electrical}.
The increasing charging load from charging stations threatens the stability of distribution network due to the high load peak, potential voltage drops, and transmission line overloads \cite{clement2009impact}. Meanwhile, high charging power is an important feature of next-generation EV charging stations \cite{engelhardt2022energy}. If the high peak charging power is provided solely by the grid, it will impose greater pressure on the grid operation. Installing energy storage can mitigate this adverse impact by providing part of the charging power during the peak hours, thereby reducing the capacity requirement on the grid. Moreover, energy storage can help accommodate the fluctuating renewable generation in charging stations \cite{ref1}, which otherwise needs to be handled by the grid. Energy storage can also simultaneously provide multiple services, such as load shifting, load balancing, and primary frequency response \cite{zhong2021chance}. Hence, it is necessary to equip charging stations with energy storage.
However, since energy storage is expensive, it may be uneconomical to deploy individual energy storage for each charging station. An alternative solution is to allow multiple charging stations to access and share a common energy storage \cite{dai2021utilization}. Applying shared energy storage is promising and will change the current architecture and operation of charging stations.
It is crucial to explore how to coordinate the EV charging stations and distribution network to adapt to the new shared energy storage architecture.

Recently, a vast literature has investigated the optimal operation of multiple EV charging stations.
An optimal EV charging scheduling algorithm was proposed to maximize the utilization of local photovoltaic (PV) generation \cite{chen2020blockchain}, and to align with the wind generation of charging stations \cite{yang2018distributed}. 
Due to the different EV charging demand patterns and renewable generations, charging stations may present distinct supply and demand characteristics.
To improve their operational efficiency, reference \cite{Affolabi2022optimal} proposed a centralized framework to coordinate the energy trading among a group of EV charging stations.
A bilevel optimization model was proposed to settle  the EV charging scheduling problem of multiple charging stations considering the underlying distribution market clearing \cite{xie2019optimal}. It was then converted into a single-level optimization problem that can be solved in a centralized way.
Reference \cite{nezamabadi2020arbitrage} adopted a similar centralized manner to solve the multiple microgrids' energy trading problem.
However, the centralized model above requires complete information on charging stations, which may cause privacy concerns and high communication burdens. The centralized model also fails to characterize the competing interests of different charging stations. Therefore, a distributed coordination mechanism is desired \cite{liu2017distributed}.
A distributed hierarchical strategy was proposed in \cite{zhang2021distributed} to coordinate the distribution network and charging stations.
Moreover, literature on energy trading among prosumers \cite{chen2023energy,2017sg}, microgrids \cite{wang2018incentivizing}, and energy buildings \cite{cui2019building}, can also provide some insights into the coordination mechanism design for charging stations. As mentioned above, in future power systems, shared energy storage is expected to play an important role in mitigating the adverse impact of unpredictable charging demand. Despite the fruitful research on the coordination of charging stations, shared energy storage was rarely considered.

In recent years, with the success of sharing economy in housing and transportation, shared energy storage has attracted much attention particularly at the residential energy user side \cite{chen2022review}.
The shared energy storage model in this paper refers to a group of users connected to a common energy storage, operated by an independent energy storage operator \cite{zhao2019virtual}. Users can buy power and capacity from the shared energy storage to reduce their own energy costs.
Reference \cite{zhang2020service} proposed a community shared energy storage to serve different residential users. Thanks to the complementary load profiles of different users, the sharing business model can reduce the required capacity \cite{liu2017decision} and improve the utilization \cite{yang2021optimal} of energy storage.
To coordinate the shared energy storage operator and users, a non-cooperative game was formulated to determine the service price of shared energy storage and the purchase of users \cite{fleischhacker2018sharing}. However, a sub-optimal equilibrium may happen due to the inherent competing interests. To achieve social optimum, a centralized operation mechanism was adopted to determine the shared energy storage operation and users' charging/discharging strategies \cite{walker2021analysis}. 
To overcome the shortcomings of centralized mechanisms, reference \cite{zhu2021distributed} proposed a distributed strategy to deal with the case that multiple households invest and control a common energy storage.
In practice, sharing energy storage is more commonly owned by an independent operator rather than users. An alternating direction method of multipliers (ADMM)-based distributed algorithm was proposed in \cite{zhong2019online} to fit for this setting.
However, the studies above are based on a simplified model neglecting distribution network constraints and the charging station operation is also more complicated than a household user.

\begin{table*}[!htbp]\label{tab:refcmp}%
\centering
\caption{Comparison of this paper with other related literature}
    \begin{tabular}{cccccc}
    \toprule
    \multicolumn{1}{c}{\multirow{2}[0]{*}{Reference}} & \multicolumn{1}{c}{Network} & \multicolumn{1}{c}{\multirow{2}[0]{*}{Distributed}} & \multicolumn{1}{c}{Shared energy} & Three types of & Profit \\
          & constraints &       & storage & stakeholders & allocation \\
    \midrule
    \cite{chen2020blockchain,yang2018distributed} & x     & x     & x     & x     & x \\
    \cite{Affolabi2022optimal} --\cite{nezamabadi2020arbitrage} & $\checkmark$     & x     & x     & x     & x \\
\cite{liu2017distributed,zhang2021distributed,chen2023energy} & $\checkmark$     & $\checkmark$     & x     & x     & x \\
    \cite{2017sg} & x     & $\checkmark$     & x     & x     & x \\
    \cite{wang2018incentivizing,cui2019building} & x     & $\checkmark$     & x     & x     & $\checkmark$ \\
    \cite{chen2022review} --\cite{yang2021optimal},\cite{walker2021analysis} & x     & x     & $\checkmark$     & x     & x \\
    \cite{fleischhacker2018sharing,zhu2021distributed,zhong2019online} & x     & $\checkmark$     & $\checkmark$     & x     & x \\
    This paper & $\checkmark$     & $\checkmark$     & $\checkmark$     & $\checkmark$     & $\checkmark$ \\
    \bottomrule
    \end{tabular}%
\end{table*}%

A distributed coordination mechanism that considers both distribution network constraints and shared energy storage is not trivial. The charging stations, shared energy storage, and distribution network are operated by different agents with competing interests. The coordination mechanism should enable individual decision-making for the three different groups of agents. Though the ADMM algorithm has been widely used in designing distributed coordination mechanisms, the conventional ADMM \cite{2011boyd} can only handle the case with two groups of agents. For example, in \cite{zhang2021distributed}, one group consists of the charging stations while the other group is the DSO. More precisely, we call it a ``two-block'' ADMM. Unfortunately, it has been shown that the direct extension of ADMM to three-/multi-block has no convergence guarantee \cite{chen2016direct,he2018class}. Therefore, an innovative distributed coordination mechanism is needed to adapt to the new architecture.

We compare the EV charging station coordination methods proposed in this paper and other related literature in Table \ref{tab:refcmp}.
The main research gaps are summarized as follows:
1)	Previous work on charging station coordination mainly adopted an individual energy storage architecture. Considering the potential of shared energy storage in terms of cost reduction, the coordination of charging stations with shared energy storage is an important topic that has not been fully studied. Most existing literature on shared energy storage used a cooperative game approach \cite{yang2021optimal,chakraborty2019sharing}, which cannot well capture the conflicting interests among different agents. To fill this gap, this paper proposes an equilibrium model among charging stations, shared energy storage, and distribution networks, and demonstrates the existence of an equilibrium and the condition that the trading prices should meet.
2)	How to achieve the equilibrium is another problem worth studying. Considering the similarity between the decision-making problems of each individual agent and the distributed optimization framework, the ADMM algorithm is proposed to be used for the mechanism design. However, the direct extension of ADMM from two-block to three-block is not necessarily convergent \cite{chen2016direct}. To fill this gap, a modified ADMM algorithm with a prediction and a correction step is developed to search for the equilibrium, whose convergence is proved.

Our main contributions are two-fold:
\begin{enumerate}
  \item \emph{An equilibrium model between CSOs, SESOs, and DSO.} In this paper, an equilibrium model is developed to characterize the interaction among different agents and the condition under which the energy supplies meet the energy demands. Unlike the previous work that used cooperative game-based approaches, the proposed model takes into account the interest conflicts among different stakeholders. We prove that, if an equilibrium exists, it must be the optimal solution of the associated centralized operation problem, and that the energy trading prices equal the value of dual variables at the optimum. To the best of our knowledge, this finding has not been reported in the existing literature.
  \item \emph{Modified ADMM algorithm to search for the equilibrium.} To search for the equilibrium, a distributed coordination mechanism based on a modified ADMM algorithm is developed. Distinct from the traditional ADMM algorithm, the proposed one has an additional corrective step to ensure convergence even when dealing with models having a ``three-block'' form (e.g., the problem studied in this paper). We prove that the modified ADMM algorithm can output the desired energy trading prices that delivers the system equilibrium. Extension from the traditional ADMM algorithm to the proposed one is not trivial, which requires a completely different way to prove the convergence.
\end{enumerate}

The rest of this paper is organized as follows. Section \ref{sec:model} describes the overall system structure and models, and gives the equilibrium model. Section \ref{sec:distributed} introduces the proposed distributed algorithm with a proof of convergence. Simulation results are presented in Section \ref{sec:result}. Finally, section \ref{sec:conclu} concludes the paper.

\section{Problem Formulation}\label{sec:model}
In this section, we first provide an overview of the system framework and then model the three groups of participants, i.e., the charging station, shared energy storage, and distribution network, respectively. Then, an equilibrium model is formulated to characterize the interaction among different participants, whose property is proved theoretically.

\begin{figure}[!htbp]
  \centering
  \includegraphics[width=0.48\textwidth]{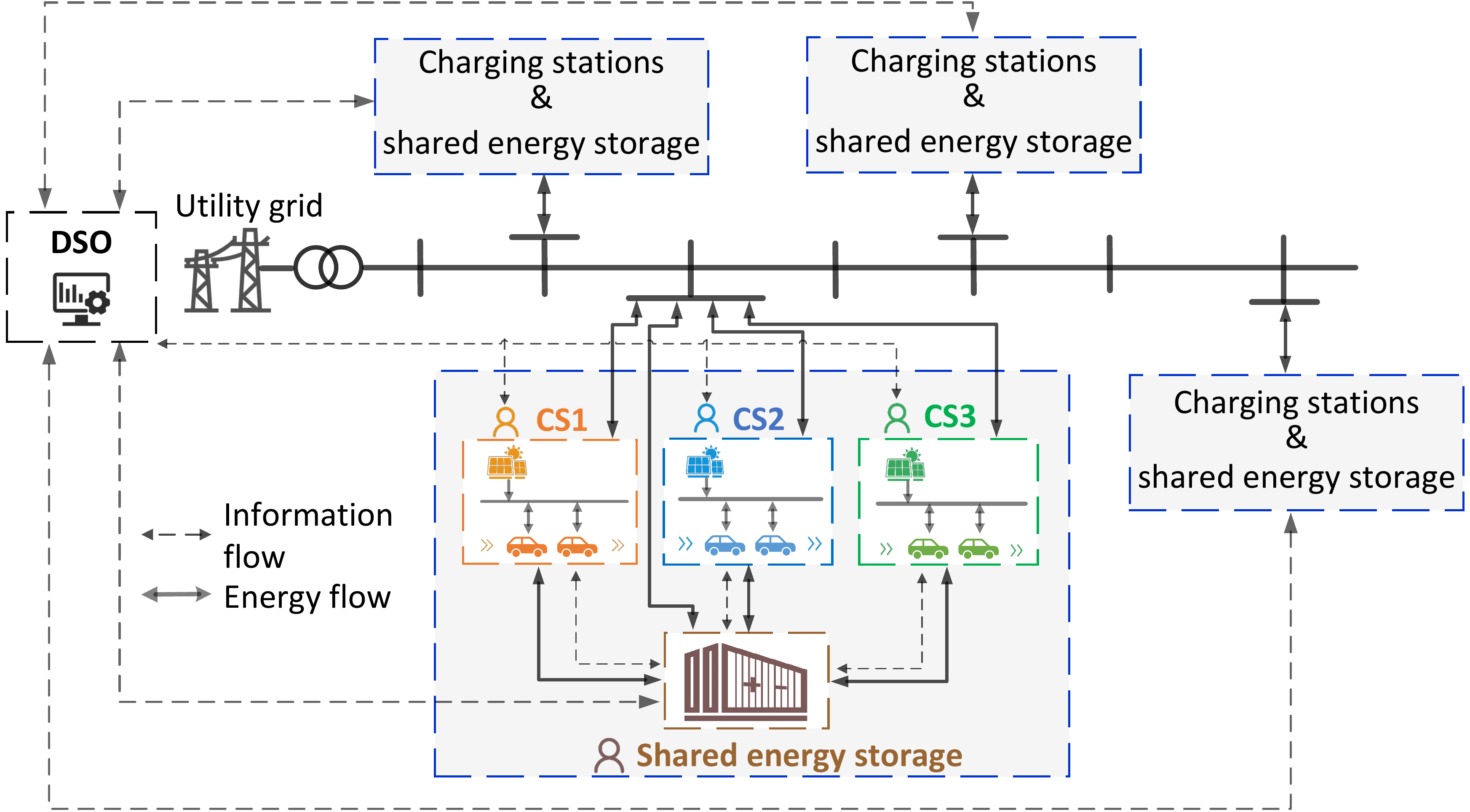}\\
  \caption{Architecture of the EV charging stations system with shared energy storage in a distribution network.}\label{fig:sysConf}
\end{figure}

\subsection{System Framework}
Fig.~\ref{fig:sysConf} shows the overall system architecture, including the EV charging stations (CS), shared energy storage (SES), and distribution network.  $\mathcal{I}$ is defined as the set of EV charging stations, and each one is indexed by $i\in\mathcal{I}$. Similarly, $\mathcal{B}$ is the set of SES, each of which is indexed by $b\in\mathcal{B}$.
We assume that the multiple charging stations that share a common energy storage are not far from each other. Therefore, as shown in Fig. \ref{fig:sysConf}, the charging stations and their shared energy storage are located at one node of a city-sized distribution network. In such a circumstance, it is reasonable to ignore the energy transmission losses between the energy storage and charging stations. A similar setting can be found in \cite{wang2018incentivizing,cui2021community}.
Let $\mathcal{I}_b$ denote the set of CSs connected to SES $b$.
The SES can provide services for the connected CSs. Thus, CSs do not need to deploy local individual energy storage, saving a large amount of investment cost.
In addition, similar to the multi-service architecture in \cite{zhong2021chance}, we allow shared energy storage to serve both the distribution network and charging stations, as shown in Fig. \ref{fig:sysConf}.
Both the CSs and SES can buy/sell electricity from/to the distribution power system.
Let $\mathcal{T}=\{1,...,T\}$ be the time horizon, and every time slot is indexed by $t\in \mathcal{T}$ with an equal time interval $\Delta t=1$ hour.

\begin{figure}[!htbp]
  \centering
  \includegraphics[width=0.2\textwidth]{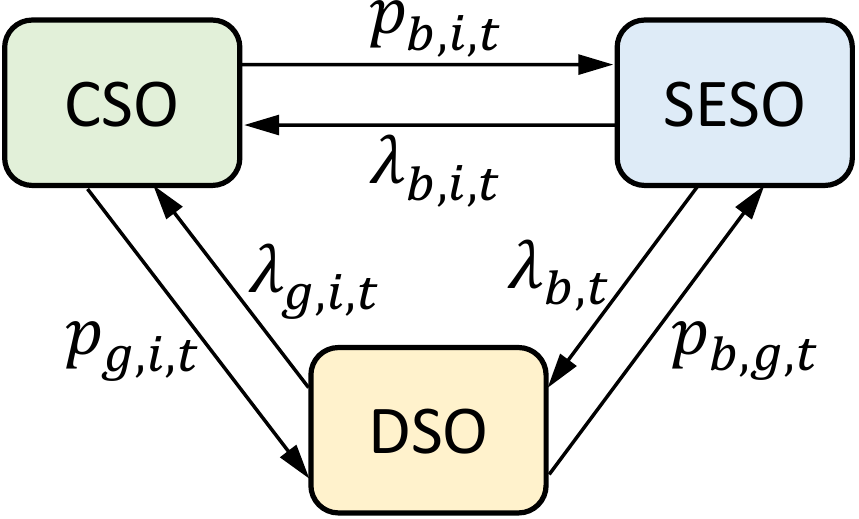}\\
  \caption{Equilibrium among CSO, SESO, and DSO.}\label{fig:equil}
\end{figure}

Despite the potential huge cost-saving, the shared energy storage architecture complicates the energy management of the overall system compared with the case with individual energy storage. To be specific, the distributed network, the charging stations, and the shared energy storage belong to three different stakeholders, i.e., the distribution network operator (DSO), charging station operator (CSO), and shared energy storage operator (SESO), with competing interests. The information possessed by these three stakeholders are asymmetric. For example, the network constraint is known only by the DSO but not the CSO and SESO. Therefore, an appropriate equilibrium model is necessary to characterize the decision-making of different stakeholders and the interaction among them. In the following, we first introduce the models of CSO, SESO, and DSO, respectively. Then, an equilibrium model is built to describe a situation when the electricity supplies and demands among them are balanced.
As seen in Fig. \ref{fig:equil}, CSO $i$ exchanges energy $p_{b,i,t}$ with SESO at a price $\lambda_{b,i,t}$, CSO $i$ exchanges energy $p_{g,i,t}$ with DSO at a price $\lambda_{g,i,t}$, and SESO exchanges energy $p_{b,g,t}$ with DSO at a price $\lambda_{b,t}$.

\subsection{EV Charging in Charging Station}
\begin{figure}[!htbp]
  \centering
  \includegraphics[width=0.4\textwidth]{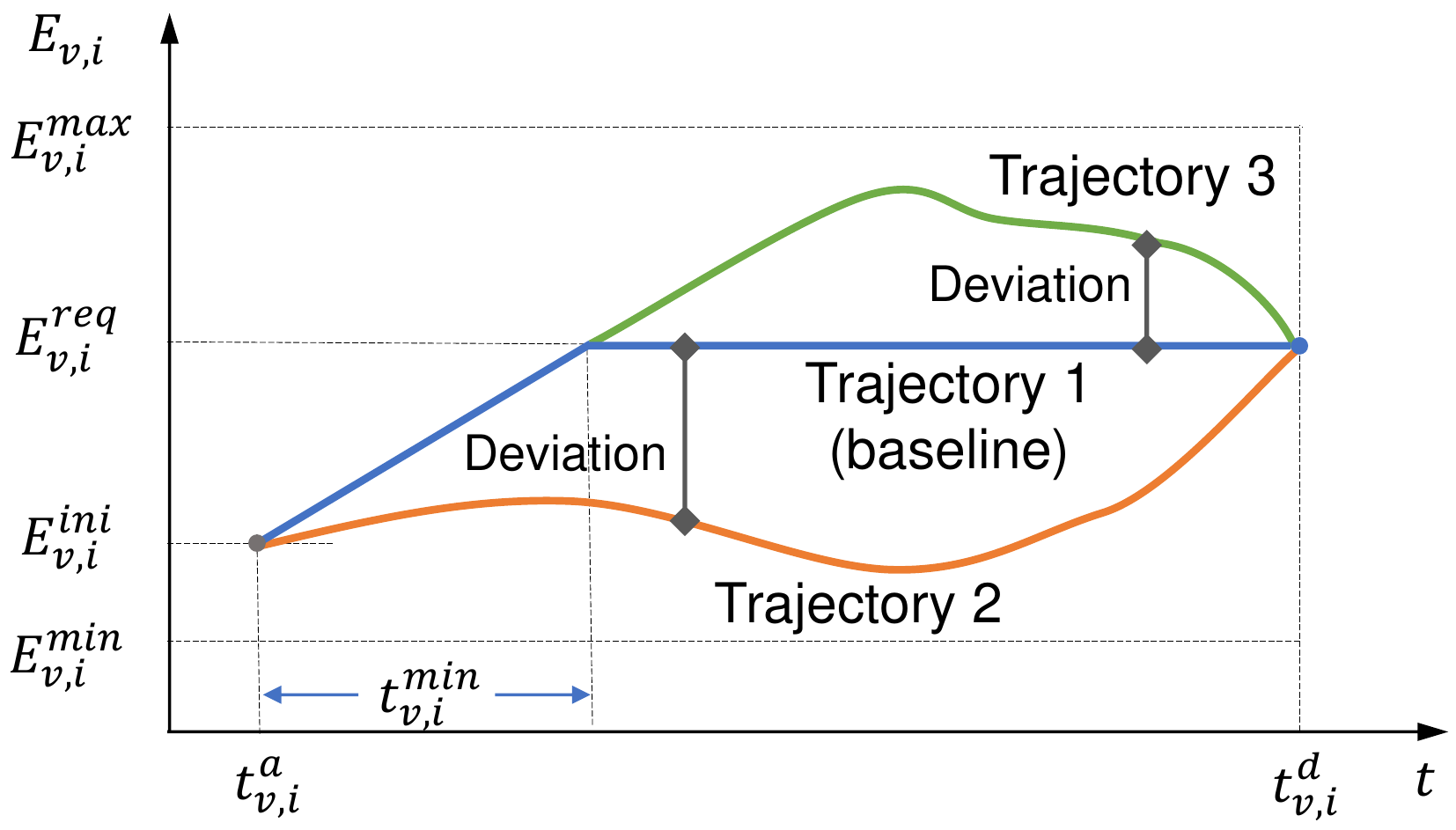}\\
  \caption{An EV's energy curves under different charging trajectories.}\label{fig:evcharging}
\end{figure}

We denote $\mathcal{V}_i$ as the set of EVs to be charged in CS $i\in\mathcal{I}$. For each EV $v\in\mathcal{V}_i$, its charging need can be specified by four parameters: $(t_{v,i}^a, t_{v,i}^d, E_{v,i}^{ini}, E_{v,i}^{req})$. Here, $t_{v,i}^a$ and $t_{v,i}^d$ represent the arrival time and anticipated departure time of EV $i$, and $E_{v,i}^{ini}$ and $E_{v,i}^{req}$ are the initial EV battery energy level and the required energy level when leaving, respectively.
This means that the EV requires its energy level to be at least $E_{v,i}^{req}$ but not necessarily the higher the better. For example, the EV owner may wish to charge its EV to a certain energy level in order to drive back home, where it can charge its EV fully at a lower cost. Trajectory 1 refers to the benchmark that the EV is charged at maximum charging power upon arrival until its charging requirement $E_{v,i}^{req}$ is met. Trajectories 2 and 3 in Fig. \ref{fig:evcharging} show two potential EV energy level changes under flexible charging.

With the increasing penetration of renewable energy in distribution networks, charging EVs flexibly becomes even more important to deal with fluctuating and intermittent power supplies. However, this may lead to dissatisfaction or inconvenience for EVs. For example, if charging according to Trajectory 1, the EV can leave at any time after $t_{v,i}^a+t_{v,i}^{min}$ with its battery energy level meeting the requirement. But if flexible charging is allowed, the EV faces the risk of not being well charged if it leaves earlier, resulting in possible inconvenience for the EV owner.
Meanwhile, charging and discharging may cause depreciation cost for EVs in CS $i$,
\begin{align}\label{equ:objcs}
    C_{cs,i}=\sum_{t\in\mathcal{T}}\sum_{v\in\mathcal{V}_i}\Big[c_{v,i}(p_{v,i,t}\Delta t-\bar{p}_{v,i,t}\Delta t)^2\nonumber\\
    +c_{v,i}^{dep}(p_{v,i,t}^{c}\Delta t+p_{v,i,t}^{d}\Delta t)\Big],
\end{align}
where $c_{v,i}$ is the cost coefficient to measure the inconvenience of EV $v$ \cite{cui2021community}, $p_{v,i,t}$ is the actual charging power of EV $v$ at time $t$, $\bar{p}_{v,i,t}$ is the EV owner's desired charging power (charging as soon as possible, CASAP), $c_{v,i}^{dep}$ represents the coefficient used to measure the depreciation cost associated with EV charging and discharging \cite{wang2018incentivizing}, $p_{v,i,t}^c$ and $p_{v,i,t}^d$ are the charging and discharging power of EV $v$ in CS $i$.
The first quadratic term measures the deviation between the actual charging power trajectory $p_{v,i,t},\forall t$ (e.g., Trajectory 2 or 3) and the EV owner's desired charging power trajectory $\bar{p}_{v,i,t},\forall t$ (e.g., Trajectory 1). The second term measures the depreciation cost due to charging and discharging.
If we let $E_{v,i}^{cha}=E_{v,i}^{req}-E_{v,i}^{ini}$, then $\bar{p}_{v,i,t}$ can be determined by
\begin{gather}\label{equ:evbaseline}\small
    \bar{p}_{v,i,t}=\left\{
    \begin{array}{ll}
    p_{v,i}^{max},     & t_{v,i}^a\leq t<\lfloor t_{v,i}^{min}\rfloor+t_{v,i}^a \\
    E_{v,i}^{cha}/\eta_{c,v}-\lfloor t_{v,i}^{min}\rfloor p_{v,i}^{max},     & t=\lfloor t_{v,i}^{min}\rfloor+t_{v,i}^a \\
    0,           & \text{otherwise}
    \end{array}\right.
\end{gather}
The explanation of \eqref{equ:evbaseline} is: since the charging power $\bar{p}_{v,i,t}$ corresponds to the CASAP way, during time slots $t_{v,i}^a\leq t<t_{v,i}^a+\lfloor t_{v,i}^{min}\rfloor$, EV uses the maximum charging power $p_{v,i}^{max}$, where $t_{v,i}^{min}=(E_{v,i}^{req}-E_{v,i}^{ini})/(p_{v,i}^{max} \eta_{c,v})$ means the minimum time needed for EV $v$ in CS $i$ to be charged to its required energy level $E_{v,i}^{req}$, $\eta_{c,v}$ is the charging efficiency, and $\lfloor.\rfloor$ means rounding down to the nearest integer. For example, if $E_{v,i}^{req}=30$, $E_{v,i}^{ini}=10$, $p_{v,i}^{max}=6.6$, $\eta_{c,v}=0.95$, then $t_{v,i}^{min}=\frac{30-10}{6.6\times 0.95}=3.19$h and $\lfloor t_{v,i}^{min}\rfloor=3$h. Then, at the next time slot $t=t_{v,i}^a+\lfloor t_{v,i}^{min}\rfloor$, the EV will be charged with $\bar{p}_{v,i,t}=E_{v,i}^{cha}/\eta_{c,v}-\lfloor t_{v,i}^{min}\rfloor p_{v,i}^{max}$ so that its final energy level will equal what it declares. Afterwards, since the charging task has been fulfilled, the charging power becomes zero.

Other constraints that EV charging should satisfy are
\begin{subequations}\label{eq:EV}
\begin{gather}
  0\leq p_{v,i,t}^{c}\leq p_{v,i}^{max},\forall v,\forall t \in [t_{v,i}^a, t_{v,i}^d],\label{equ:pc}\\
  0\leq p_{v,i,t}^{d}\leq p_{v,i}^{max},\forall v,\forall t \in [t_{v,i}^a, t_{v,i}^d],\label{equ:pd}\\
  p_{v,i,t}=p_{v,i,t}^{c}-p_{v,i,t}^{d},\forall v,\forall t,\label{equ:pcd}\\
  p_{v,i,t}^c=0,~ p_{v,i,t}^d=0,~\forall v,\forall t \notin [t_{v,i}^a, t_{v,i}^d],\label{equ:time}\\
  E_{v,i,t+1}= E_{v,i,t}+p_{v,i,t}^{c}\Delta t\eta_{c,v}-p_{v,i,t}^{d}\Delta t/\eta_{d,v},\forall v,\forall t, \label{equ:Ev}\\
  E^{min}_{v,i}\leq E_{v,i,t}\leq E^{max}_{v,i},\forall v,\forall t, \label{equ:Evranges}\\
  E_{v,i,t_v^a}= E_{v,i}^{ini},\forall v,\label{equ:Evta}\\
  E_{v,i,t_v^d}= E_{v,i}^{req},\forall v.\label{equ:Evtd}
\end{gather}
\end{subequations}
Constraints (\ref{equ:pc}) and (\ref{equ:pd}) limit the charging/discharging power of an EV $v$ during its available time interval $[t_{v,i}^a, t_{v,i}^d]$, where $p_{v,i}^{\max}$ is the maximum charging/discharging power.
Constraint (\ref{equ:pcd}) defines the actual charging power of an EV $v$.
Constraint (\ref{equ:time}) sets the charging/discharging power of EV $v$ to zero when it is not available for charging.
Constraint (\ref{equ:Ev}) describes the EV's energy dynamics, where $\eta_{c,v}/\eta_{d,v}$ is the charging/discharging efficiency.
Constraint (\ref{equ:Evranges}) ensures that the EV's energy will not be lower than the minimal energy $E^{min}_{v,i}$ nor exceed the maximal energy $E^{max}_{v,i}$.
Constraint (\ref{equ:Evta}) gives the EV's initial energy.
Constraint (\ref{equ:Evtd}) ensures the satisfaction of charging requirements when the EVs depart.

Let
\begin{align} \label{eq:EVaggregate}
    p_{d,i,t}=\sum_{v \in \mathcal{V}_i} p_{v,i,t}
\end{align}

The CS $i \in \mathcal{I}$ can use its on-site PV generation, whose power output is $p_{pv,i,t}$, to serve its charging demand. If there is surplus electricity, it can sell $p_{g,i,t}$ to the distribution network at a price $\lambda_{g,i,t}$ or sell $p_{b,i,t}$ to the SES at a price $\lambda_{b,i,t}$, respectively. The power balancing condition for the CS $i$ is
\begin{align}\label{eq:CSO-balance}
    p_{d,i,t}+p_{g,i,t}+p_{b,i,t}=p_{pv,i,t},\forall i,\forall t.
\end{align}

The CSO aims to minimize its overall operation cost by solving:
\begin{subequations}\label{eq:CSO}
\begin{align}
    \min ~ & C_{cso,i}=C_{cs,i}-\sum_{t \in \mathcal{T}} \lambda_{g,i,t}p_{g,i,t} - \sum_{t \in \mathcal{T}} \lambda_{b,i,t} p_{b,i,t}, \\
    \mbox{s.t.}~ & \eqref{eq:EV}-\eqref{eq:CSO-balance}.
\end{align}
\end{subequations}

\subsection{Shared Energy Storage Operation}
A SES $b\in\mathcal{B}$ provides energy storage services to both its interconnected CS $i\in\mathcal{I}_b$ and the distribution network.
The resulting charging and discharging can lead to energy storage's degradation cost
\begin{align}
C_{bat,b}(p^d_{b,i,t},p^c_{b,i,t})= \sum\limits_{i\in\mathcal{I}_b}\sum\limits_{t\in\mathcal{T}}c_{b}\left(p^d_{b,i,t}+p^c_{b,i,t}\right)\Delta  t \nonumber\\
+ \sum_{t \in \mathcal{T}}c_b (p_{b,g,t}^d+p_{b,g,t}^c)\Delta t,\label{equ:batcost}
\end{align}
where $p^{d}_{b,i,t}$ and $p^{c}_{b,i,t}$ are the discharging and charging power provided by the SES $b$ to CS $i\in\mathcal{I}_b$ at time slot $t\in\mathcal{T}$; $p_{b,g,t}^d$, $p_{b,g,t}^c$ are the discharging and charging power provided to the grid. $c_{b}$ is the degradation cost coefficient of SES $b\in\mathcal{B}$ \cite{wang2018incentivizing}.
The $p^{d}_{b,i,t}, p^{c}_{b,i,t}$ and $p_{b,g,t}^d$, $p_{b,g,t}^c$ should meet the following physical constraints:
\begin{gather}
  0\leq \sum_{i\in\mathcal{I}_b}p^{d}_{b,i,t}+p_{b,g,t}^d\leq p_{b}^{d,max},\label{equ:pbdch}\\
  0\leq \sum_{i\in\mathcal{I}_b}p^{c}_{b,i,t}+p_{b,g,t}^c\leq p_{b}^{c,max},\label{equ:pbcha}\\
  p^{c}_{b,i,t}\geq 0,  p^{d}_{b,i,t}\geq 0,\forall i \in \mathcal{I}_b,
\end{gather}
where $p_{b}^{d,max}$ is the maximum discharging power and $p_{b}^{c,max}$ is the maximum charging power. Constraint (\ref{equ:pbdch}) means that the sum of discharging power from all interconnected CSs cannot exceed the maximum discharging power of SES at any time. The same is true for charging power in constraint (\ref{equ:pbcha}). Let $p_{b,i,t}$ and $p_{b,g,t}$ be the power that CS $i$ and distribution network deliver to SES $b$, respectively, then
\begin{align}
p_{b,i,t}=~ & p_{b,i,t}^c-p_{b,i,t}^d, \label{eq:battery}\\
p_{b,g,t}=~ & p_{b,g,t}^c-p_{b,g,t}^d. \label{eq:battery-2}
\end{align}

Along with discharging and charging, the energy dynamics of shared energy storage can be expressed by:
\begin{align}\label{equ:Ebat}
E_{b,t+1}= E_{b,t}-(\sum\limits_{i\in\mathcal{I}_b}p^{d}_{b,i,t}+p_{b,g,t}^d)\Delta t/\eta_d \nonumber\\
 +(\sum\limits_{i\in\mathcal{I}_b}p^{c}_{b,i,t}+p_{b,g,t}^c)\Delta t\eta_c,
\end{align}
where $E_{b,t}$ represents the remained energy of shared energy storage at time $t$; $\eta_d$/$\eta_c$ is the discharging/charging efficiency.
$E_{b,t}$ should always be within its allowable range to ensure no over-discharging nor over-charging occurs, i.e.,
\begin{equation}\label{equ:EbatInequ}
E^{min}_{b}\leq E_{b,t}\leq E^{max}_{b},
\end{equation}
where $E_{b}^{min}$ and $E_{b}^{max}$ are the minimal and maximal allowable energy levels of the shared energy storage.

Besides, the energy level of SES at the initial ($t=1$) and final ($t=T$) time slots are restricted to be equal.
\begin{equation}\label{equ:soct1tT}
E_{b,1}= E_{b,T}.
\end{equation}
This restriction is commonly used in references such as \cite{wang2018incentivizing,qiu2018bi,li2020risk} and is based on real scenes. It requires that the energy storage's initial and ending energy levels are identical for each dispatch cycle. This enables the decoupling of energy storage operation in different dispatch cycles.

Suppose the electricity price at which the shared energy storage buys from the distribution network is $\lambda_{b,t}$. Then, Each SESO solves:
\begin{subequations}\label{eq:SESO}
\begin{align}
 \min~ & C_{seso,b}=C_{bat,b} + \sum_{t \in \mathcal{T}} \sum_{i \in \mathcal{I}_b} \lambda_{b,i,t} p_{b,i,t} + \sum_{t \in \mathcal{T}} \lambda_{b,t} p_{b,g,t}, \\
 \mbox{s.t.}~ & \eqref{equ:pbdch}-\eqref{equ:soct1tT}.
\end{align}
\end{subequations}
where the second and third terms are the trading costs with the CSOs and DSO, respectively. A negative $C_{seso,b}$ means that SESO obtains profit, i.e., its revenue from selling electricity to the CSOs and DSO is greater than its operational cost.

\subsection{Distribution Network Operation}
We consider a radial distribution network with EV charging stations located at its different buses. To fully capture the impact of charging stations' operation on the electric grid, the distribution network model should be considered.
The distribution network can be represented by a graph $\mathcal{G(N,E)}$, where $\mathcal{N}$ is the set of buses and $\mathcal{E}$ is the set of lines.
We index each bus in $\mathcal{N}$ by $n=1, 2, ... ,N$, and each branch by $(n,j)\in\mathcal{E}$, meaning that the line directs from $n$ to $j$, where node $n$ is closer to the root than node $j$.
The branch flow model is adopted \cite{low2013branch}, $\forall t$:
\begin{subequations}
\label{eq:ACpowerflow}
\begin{align}
p_{j,t}=& ~ P_{nj,t}-r_{nj}\ell_{nj,t}-\!\!\sum_{k:(j,k)\in\mathcal{E}}P_{jk,t},\forall (n,j) \in \mathcal{E},\label{equ:pinpf}\\
q_{j,t}=&~Q_{nj,t}-x_{nj}\ell_{nj,t}-\!\!\sum_{k:(j,k)\in\mathcal{E}}Q_{jk,t},\forall (n,j) \in \mathcal{E},\\
v_{j,t}=&~ v_{n,t}-2(r_{nj}P_{nj,t}+x_{nj}Q_{nj,t})\nonumber\\
~& +(r^2_{nj}+x^2_{nj})\ell_{nj,t}, \forall (n,j) \in \mathcal{E}, \label{equ:pfvj}\\
\ell_{nj,t}= & \frac{P_{nj,t}^2+Q_{nj,t}^2}{v_{n,t}},\forall (n,j) \in \mathcal{E},\label{equ:pfnonconvex} \\
\underline{p_j}\leq &~ p_{j,t}\leq \overline{p_j},~\underline{q_j}\leq q_{j,t}\leq \overline{q_j},\forall j \in \mathcal{N},\label{equ:pfpqub}\\
0\leq&~ \ell_{nj,t}\leq \overline{\ell_{nj}},~\forall (n,j) \in \mathcal{E},\\
\underline{v_j}\leq &~ v_{j,t}\leq \overline{v_j},\forall j \in \mathcal{N},\label{equ:pflvub}
\end{align}
\end{subequations}
where $p_{j,t}/q_{j,t}$ is the active/reactive power of bus $j\in\mathcal{N}$ at time slot $t$, $P_{nj,t}/Q_{nj,t}$ is the active/reactive power flow of line $(n,j)$, $r_{nj}/x_{nj}$ is the resistance/reactance of line $(n,j)$, $\ell_{nj,t}$ is the squared current magnitude of line $(n,j)$, $v_{j,t}$ is the squared voltage magnitude at bus $j$, $\underline{\bullet}$ and $\overline{\bullet}$ represent the lower and upper bounds of the variable $\bullet$, respectively. The constraint (\ref{equ:pfnonconvex}) is nonconvex, which can be relaxed into a second-order cone inequality constraint, i.e.,
\begin{equation}\label{equ:socp}
\ell_{nj,t}\geq\frac{P_{nj,t}^2+Q_{nj,t}^2}{v_{n,t}},\forall (n,j) \in \mathcal{E}.
\end{equation}
For the radial distribution network we considered in this paper, under some mild conditions (e.g., the bus voltages are around the nominal value), the above relaxation is exact \cite{low2013branch}.

Suppose $n_i$ is the bus to which the SES $b \in \mathcal{B}$ and CS $i \in \mathcal{I}_b$ are connected. Then, we have
\begin{equation}\label{equ:pni=pgi}
    p_{n_i,t} + \sum_{i \in \mathcal{I}_b} p_{g,i,t} + p_{g,b,t}= 0, 
\end{equation}
where $p_{g,b,t}$ is the purchased power by the distribution network from SES $b$. Hence,
\begin{align}\label{eq:DSO-balance}
    p_{g,b,t}+p_{b,g,t}=0,\forall t.
\end{align}

The DSO is responsible for the distribution network management and buys/sells from/to the utility grid for power balance.
Suppose bus 1 in the distribution network is the slack bus connected to the utility grid, and the DSO aims to minimize its cost for electricity buying and selling, i.e.,
\begin{equation}\label{equ:objdso}
    \min~C_{ds}(p_{j,t})
    =\sum_{t \in \mathcal{T}}(p^b_{1,t}\lambda^b_{g,t}-p^s_{1,t}\lambda^s_{g,t}),
\end{equation}
where $p_{1,t}^b/p_{1,t}^s$ is the energy bought/sold from/to the utility grid by bus 1 at time slot $t$. $\lambda^b_{g,t}/\lambda^s_{g,t}$ is the utility electricity purchase/sale prices. In addition, $\lambda^s_{g,t}<\lambda^b_{g,t}$ should be satisfied to avoid DSO arbitrage by simultaneously buying from and selling back to the utility grid. 
The DSO minimizes its overall operational cost by solving:
\begin{subequations}\label{eq:DSO}
\begin{align}
    \min ~ & C_{dso}=C_{ds} + \sum_{t \in \mathcal{T}} \sum_{i \in \mathcal{I}} \lambda_{g,i,t} p_{g,i,t} -\sum_{t \in \mathcal{T}} \sum_{b \in \mathcal{B}} \lambda_{b,t} p_{b,g,t} \\
    \mbox{s.t.}~ & \eqref{equ:pinpf}-\eqref{equ:pfvj}, \eqref{equ:pfpqub}-\eqref{equ:pflvub}, \eqref{equ:socp}-\eqref{eq:DSO-balance}
\end{align}
\end{subequations}

\subsection{Equilibrium Model}
Given the prices $(\lambda_{g,i,t},\forall i,\forall t; \lambda_{b,t},\forall b,\forall t; \lambda_{b,i,t},\forall i,\forall t)$, the amount of electricity CSOs, SESOs, and DSO want to sell/buy to/from each other can be determined by solving the problems \eqref{eq:CSO}, \eqref{eq:SESO}, and \eqref{eq:DSO}, respectively. The equilibrium of the overall system is defined as follows:

\begin{definition} (System Equilibrium, SE)
Given the price $\lambda:=(\lambda_{g,i,t},\forall i \in \mathcal{I},\forall t; \lambda_{b,t},\forall b \in \mathcal{B},\forall t; \lambda_{b,i,t},\forall i \in \mathcal{I},\forall t)$, let $(p_{g,i,t}^*(\lambda), \forall t; p_{b,i,t}^*(\lambda),\forall t)$ be the optimal solution of \eqref{eq:CSO} for all $i \in \mathcal{I}$; let $(p_{b,i,t}^{'}(\lambda), \forall i \in \mathcal{I}_b,\forall t; p_{b,g,t}^{*}(\lambda),\forall t)$ be the optimal solution of \eqref{eq:SESO} for all $b \in \mathcal{B}$; let $(p_{g,i,t}^{'}(\lambda), \forall i \in \mathcal{I},\forall t; p_{b,g,t}^{'}(\lambda),\forall b\in \mathcal{B}, \forall t)$ be the optimal solution of \eqref{eq:DSO}. Then, the system equilibrium is reached if
\begin{subequations}\label{eq:supply-demand}
\begin{align}
    p_{g,i,t}^*(\lambda)=~ & p_{g,i,t}^{'}(\lambda),\forall i,\forall t, \\
    p_{b,i,t}^*(\lambda)=~ & p_{b,i,t}^{'}(\lambda),\forall i,\forall t,\\
    p_{b,g,t}^{*}(\lambda)=~ & p_{b,g,t}^{'}(\lambda),\forall b,\forall t.
\end{align}
\end{subequations}
\end{definition}

The above definition means that a system equilibrium is reached if we can find a pricing system such that the electricity supplies equal the electricity demands. First, we analyze the property of the SE by giving the following proposition. The set $\mathcal{D}_i$ collects the constraints of charging in CS $i$, i.e., \eqref{eq:EV}-\eqref{eq:EVaggregate}.
The set $\mathcal{P}$ collects the constraints of distribution network, i.e., \eqref{equ:pinpf}-\eqref{equ:pfvj}, \eqref{equ:pfpqub}-\eqref{equ:pflvub}, \eqref{equ:socp}, \eqref{equ:pni=pgi}.
The set $\mathcal{F}_b$ collects the constraints of SES $b$, i.e., (\ref{equ:pbdch})-(\ref{equ:soct1tT}).

\begin{proposition}\label{prop-1}
Let $(\bar p_{g,i,t},\forall i,\forall t; \bar p_{b,i,t},\forall i,\forall t; \bar p_{b,g,t},\forall b,\forall t)$ be the unique optimal solution of
\begin{subequations} \label{eq:central}
\begin{gather}
\min ~\sum_{i\in\mathcal{I}}C_{cs,i}+\sum_{b\in\mathcal{B}}C_{bat,b}+C_{ds},\label{equ:obj3}\\
\text{s.t.}~p_{d,i,t}+p_{g,i,t} + p_{b,i,t}= p_{pv,i,t},\label{equ:pBala}\forall i,\forall t: \lambda_{i,t},\\
p_{g,b,t}+p_{b,g,t}=0,\forall b, \forall t: \mu_{b,t},\label{eq:pgbt=pbt}\\
\{p_{d,i,t},\forall t\} \in\mathcal{D}_i,\forall i,\\
\{p_{g,i,t},\forall i, \forall t; p_{g,b,t},\forall b,\forall t\} \in\mathcal{P},\\
\{p_{b,i,t},\forall i \in \mathcal{I}_b, \forall t; p_{b,g,t},\forall t\} \in\mathcal{F}_b,\forall b.
\end{gather}
\end{subequations}\label{equ:objci}%
and $\bar \lambda_{i,t},\forall i,\forall t$ and $\bar \mu_{b,t},\forall b,\forall t$ are the values of dual variables at optimum. Then, if under the price $\lambda:=(\lambda_{g,i,t},\forall i,\forall t; \lambda_{b,t},\forall b,\forall t; \lambda_{b,i,t},\forall i,\forall t)$ a SE is reached, we have $\lambda_{g,i,t}=\lambda_{b,i,t}=\bar \lambda_{i,t},\forall i,\forall t$, $\lambda_{b,t}=\bar \mu_{b,t},\forall b,\forall t$, and
\begin{subequations}
\begin{align}
    p_{g,i,t}^*(\lambda)=~ & p_{g,i,t}^{'}(\lambda) = \bar p_{g,i,t},\forall i,\forall t, \\
    p_{b,i,t}^*(\lambda)=~ & p_{b,i,t}^{'}(\lambda) = \bar p_{b,i,t},\forall i,\forall t,\\
    p_{b,g,t}^{*}(\lambda)=~ & p_{b,g,t}^{'}(\lambda) =\bar p_{b,g,t},\forall b,\forall t.
\end{align}
\end{subequations}
\end{proposition}

The proof of Proposition \ref{prop-1} can be found in Appendix \ref{appendix-A}. Proposition \ref{prop-1} tells us that the SE always has social optimal efficiency. However, finding the prices that achieve a SE is difficult due to the complex interaction among different stakeholders. Though the prices coincide with the values of dual variables at the optimum of \eqref{eq:central}, they cannot be obtained by solving the centralized optimization directly. This is because solving the centralized optimization \eqref{eq:central} requires complete information of each charging station, shared energy storage, and distribution network. This will jeopardize participants' privacy and cause high computational and communication burdens. Hence, an important question is whether the SE is implementable, i.e., whether we can find a coordination mechanism that delivers the equilibrium.

\section{Distributed Coordination Mechanism}\label{sec:distributed}
In this section, we aim to provide a distributed coordination mechanism to reach the SE which is social optimal. Noticing that the prices under a SE equal the value of dual variables at the optimum of \eqref{eq:central}, a straightforward way is to use the distributed optimization algorithms to design the coordination mechanism. Among those distributed optimization algorithms, ADMM is the most widely-used one. However, the conventional ADMM method (or the so-called ``two-block ADMM") may not work well because the problem studied has a ``three-block form" due to the three groups of agents.
The direct extension of ADMM from two-block to three-block is not necessarily convergent \cite{chen2016direct}. The concern about possible failure in convergence would lead to distrust in the market and affect the participants' willingness to join.

To overcome this problem, we propose a distributed coordination mechanism with a convergence guarantee. It consists of a \emph{prediction} step and a \emph{correction} step. The prediction step inherits the ADMM output, and the correction step corrects the output of the prediction step to ensure convergence.

\subsection{Prediction Step}
In the prediction step, we first formulate the augmented Lagrangian function based on the coupled constraints \eqref{equ:pBala} and \eqref{eq:pgbt=pbt}, which couple the three groups of stakeholders (i.e., CSOs, SESOs, and the DSO).
\begin{equation}
    \begin{aligned}
        \mathcal{L}=&\sum_{i\in\mathcal{I}}C_{cs,i}+\sum_{b\in\mathcal{B}}C_{bat,b}+C_{ds}\\
        &-\sum_{t\in\mathcal{T}}\sum_{i\in\mathcal{I}}\lambda_{i,t}(p_{d,i,t}+p_{g,i,t} + p_{b,i,t}-p_{pv,i,t})\\
        &-\sum_{t\in\mathcal{T}}\sum_{b\in\mathcal{B}}\mu_{b,t}(p_{g,b,t}+p_{b,g,t})\\
        &+\frac{\beta}{2}\sum_{t\in\mathcal{T}}\sum_{i\in\mathcal{I}}(p_{d,i,t}+p_{g,i,t} + p_{b,i,t}-p_{pv,i,t})^2\\
        &+\frac{\beta}{2}\sum_{t\in\mathcal{T}}\sum_{b\in\mathcal{B}}(p_{g,b,t}+p_{b,g,t})^2,
    \end{aligned}
\end{equation}
where $\lambda_{i,t}$ is the dual variable for constraint \eqref{equ:pBala}, and $\mu_{b,t}$ is the dual variable for constraint \eqref{eq:pgbt=pbt}.

Based on the above Lagrangian function and referring to the ADMM iterative update steps, we independently update the decisions of CSO, SESO, and DSO in an iterative way:

\begin{subequations}
For each CSO $i\in\mathcal{I}$, it updates its own variables by solving the local optimization problem
\begin{equation}\label{equ:pditbar}
\begin{aligned}
    \min~&C_{cs,i}-\sum_{t\in\mathcal{T}} \lambda_{i,t}^k p_{d,i,t}\\
    &+\frac{\beta}{2}\sum_{t\in\mathcal{T}}\left(p_{d,i,t}+p_{g,i,t}^k + p_{b,i,t}^k - p_{pv,i,t}\right)^2\\
    \text{s.t.}~& \{p_{d,i,t},\forall t\}\in\mathcal{D}_i.
\end{aligned}
\end{equation}
We denote the solution variable as $\tilde{p}_{d,i,t}^{k}$, which is then sent to SESO and DSO for continue updating.

For each SESO $b\in\mathcal{B}$, it solves the local optimization problem
\begin{equation}\label{equ:pbitbar}
\begin{aligned}
    \min~&C_{bat,b}-\sum_{t\in\mathcal{T}}{\sum_{i\in\mathcal{I}_b}} \lambda_{i,t}^k p_{b,i,t}-\sum_{t\in\mathcal{T}}\mu_{b,t}^k p_{b,g,t}\\
    &+\frac{\beta}{2}\sum_{t\in\mathcal{T}}{\sum_{i\in\mathcal{I}_b}}\left(\tilde{p}_{d,i,t}^{k}+p_{g,i,t}^{k} + p_{b,i,t} - p_{pv,i,t}\right)^2\\
    &+\frac{\beta}{2}\sum_{t\in\mathcal{T}}\left(p_{g,b,t}^{k} + p_{b,g,t}\right)^2\\
    \text{s.t.}~& \{p_{b,i,t},\forall i \in \mathcal{I}_b, \forall t; p_{b,g,t},\forall t\} \in\mathcal{F}_b,
\end{aligned}
\end{equation}
where $\mathcal{I}_b$ refers to the set of CSs connected to SES $b$.
The optimal solutions are denoted as $\tilde{p}_{b,i,t}^{k}$ and $\tilde{p}_{b,g,t}^{k}$, which are then sent to the CSOs and DSO, respectively.

For DSO, it first solves the power flow problem
\begin{equation}\label{equ:pgitbar}
\begin{aligned}
    \min~&C_{dso}-\sum_{t\in\mathcal{T}}\sum_{i\in\mathcal{I}}\lambda_{i,t}^kp_{g,i,t}-\sum_{t\in\mathcal{T}}\sum_{b\in\mathcal{B}}\mu_{b,t}^k p_{g,b,t}\\
    &+\frac{\beta}{2}\sum_{t \in \mathcal{T}}\sum_{i\in \mathcal{I}}\left(\tilde{p}_{d,i,t}^{k}+p_{g,i,t} + \tilde{p}_{b,i,t}^k - p_{pv,i,t}\right)^2\\
    &+\frac{\beta}{2}\sum_{t\in\mathcal{T}}\left(p_{g,b,t} + \tilde{p}_{b,g,t}^{k}\right)^2\\
    \text{s.t.}~&\{p_{g,i,t},\forall i,\forall t; p_{g,b,t},\forall b,\forall t\} \in\mathcal{P}.
\end{aligned}
\end{equation}
The optimal solutions are denoted as $\tilde{p}_{g,i,t}^{k}$ and $\tilde{p}_{g,b,t}$.

Then, based on the updated $\tilde{p}_{d,i,t}^{k},\tilde{p}_{b,i,t}^{k}$, and $\tilde{p}_{g,i,t}^{k}$, the DSO further updates the dual variable $\tilde{\lambda}_{i,t}^{k}$,
\begin{equation}\label{equ:lambdabar}
    \tilde{\lambda}_{i,t}^{k} = \lambda_{i,t}^{k}- \beta(\tilde{p}_{d,i,t}^{k}+\tilde{p}_{g,i,t}^{k} + \tilde{p}_{b,i,t}^{k} - p_{pv,i,t}),\forall i,\forall t
\end{equation}

Meanwhile, based on the updated $\tilde{p}_{b,g,t}^{k}$ and $\tilde{p}_{g,b,t}^{k}$, the DSO updates the dual variable  $\tilde{\mu}_{b,t}^{k}$,
\begin{equation}\label{equ:mubar}
    \tilde{\mu}_{b,t}^{k} = \mu_{b,t}^{k}- \beta(\tilde{p}_{g,b,t}^{k}+\tilde{p}_{b,g,t}^{k}),\forall b,\forall t
\end{equation}
\end{subequations}

So far, we have completed the prediction step, but it may not have a convergence guarantee \cite{he2018class}.

\subsection{Correction Step}

To ensure convergence, the correction step is added to correct the output of the prediction step, i.e., $(\tilde{p}^{k}_{d,i,t},\tilde{p}^{k}_{b,i,t},\tilde{p}^{k}_{b,g,t},\tilde{p}^{k}_{g,i,t},\tilde{p}^{k}_{g,b,t},\tilde{\lambda}^{k}_{i,t},\tilde{\mu}^{k}_{b,t})$.
The correction step for each stakeholder is performed as follows:

\begin{subequations}
Each CSO $i \in \mathcal{I}$ updates EV charging demand $p^{k+1}_{d,i,t}$
\begin{align}
    p^{k+1}_{d,i,t} = \tilde{p}_{d,i,t}^k, \forall t.\label{equ:pditk+1}
\end{align}

Each SESO $b \in \mathcal{B}$ updates the trading power with CSOs $p^{k+1}_{b,i,t}$ and the trading power with DSO $p^{k+1}_{b,g,t}$,
\begin{align}
    p^{k+1}_{b,i,t} = &p_{b,i,t}^k-\alpha\Bigl(p^{k}_{b,i,t}-\tilde{p}^{k}_{b,i,t}\nonumber \\
           &-(1-\tau)(p^{k}_{g,i,t}-\tilde{p}^{k}_{g,i,t})\Bigr),\forall i \in \mathcal{I}_b,\forall t,\label{equ:pbitk+1}\\
    p^{k+1}_{b,g,t}=& p^{k}_{b,g,t}-\alpha\Bigl(p^{k}_{b,g,t}-\tilde{p}^{k}_{b,g,t}\nonumber \\
           &-(1-\tau)(p^{k}_{g,b,t}-\tilde{p}^{k}_{g,b,t})\Bigr),\forall t.\label{equ:pbhatk+1}
\end{align}

DSO updates the trading power with CSOs $p^{k+1}_{g,i,t}$, the trading power with SESO $p^{k+1}_{g,b,t}$, as well as the dual variables $\lambda_{i,t}^{k+1}$ and $\mu_{b,t}^{k+1}$,
\begin{align}
    p^{k+1}_{g,i,t} = &p_{g,i,t}^k-\alpha\Bigl(p^{k}_{g,i,t}-\tilde{p}^{k}_{g,i,t} \nonumber\\                                             &+\tau(p^{k}_{b,i,t}-\tilde{p}^{k}_{b,i,t})\Bigr),\forall i,\forall t, \label{equ:pgitk+1}\\
    p^{k+1}_{g,b,t} = &p_{g,b,t}^k-\alpha\Bigl(p^{k}_{g,b,t}-\tilde{p}^{k}_{g,b,t} \nonumber\\                                             &+\tau({p}^{k}_{b,g,t}-\tilde{p}^{k}_{b,g,t})\Bigr),\forall b,\forall t, \label{equ:pgbtk+1}\\
    \lambda_{i,t}^{k+1} = &\lambda_{i,t}^k-\alpha(\lambda_{i,t}^{k}-\tilde{\lambda}_{i,t}^{k}),\forall i,\forall t,\label{equ:lbditk+1}\\
    \mu_{b,t}^{k+1} = &\mu_{b,t}^k-\alpha(\mu_{b,t}^{k}-\tilde{\mu}_{b,t}^{k}),\forall b,\forall t,\label{equ:mubtk+1}
\end{align}
\end{subequations}
where $\tau$ and $\alpha$ are parameters. We will show later in Theorem \ref{thm-1} how to determine their values.

As seen, the interaction between CSOs, SESOs, and DSO happens iteratively in both the prediction step and the correction step. The iteration stops when the gaps satisfy the following condition,
\begin{equation}\label{equ:r}
  \|\boldsymbol{\lambda}^{k+1}-\boldsymbol{\lambda}^{k}\|\leq\delta, \|\boldsymbol{\mu}^{k+1}-\boldsymbol{\mu}^{k}\|\leq\delta,
\end{equation}
where $\delta$ is the error tolerance.

\subsection{Overall Algorithm}

A completed description of the proposed coordination mechanism is shown in Algorithm \ref{algo:dc}.
Under the proposed distributed coordination mechanism, each individual makes decisions based on its own constraints. For example, the EV information is only known to the CSO, the distribution network parameters are available only to the DSO, etc. Therefore, their information privacy is \emph{partially} protected \emph{to some extent}. Since the proposed mechanism still requires information exchange among different agents, and an opponent might be able to reverse-engineer the sensitive data from what we have exchanged, a distributed coordination mechanism with a full privacy guarantee will be our future work.

\begin{algorithm}[htbp]
\caption{Distributed Coordination Mechanism}\label{algo:dc}
\begin{algorithmic}[1]
\STATE{Set iteration index $k=0$, convergence error tolerance $\delta>0$, penalty parameter $\beta>0$, parameters $\alpha, \tau$.}
\STATE{DSO sets $\lambda^k_{i,t}=0,\forall i,\forall t$ and $p^k_{g,i,t}=0,\forall i,\forall t,$ for charging stations. SESOs initialize the discharging/charging power $p_{b,i,t}^k=0,\forall i,\forall t,$ for charging stations.}
\REPEAT
\STATE \textit{\textbf{Prediction Step:}}
\FOR{Each charging station $i\in\mathcal{I}$}
\STATE{CSO updates $\tilde{p}_{d,i,t}^k, \forall t,$ according to (\ref{equ:pditbar}), and sends them to DSO and SESO.}
\ENDFOR
\FOR{Each shared energy storage $b\in\mathcal{B}$}
\STATE SESO updates $\tilde{p}_{b,i,t}^k,\forall i \in \mathcal{I}_b,\forall t$ and $\tilde{p}_{b,g,t}^k, \forall t$, according to \eqref{equ:pbitbar}, and sends them to CSOs and DSO.
\ENDFOR
\STATE DSO then updates $\tilde{p}_{g,i,t}^k,\forall i,\forall t$ and $\tilde{p}_{g,b,t}^k,\forall b,\forall t$, according to \eqref{equ:pgitbar}, and updates the multipliers $\tilde \lambda_{i,t}^k,\forall i,\forall t$ and $\mu_{b,t}^k,\forall b,\forall t,$ via (\ref{equ:lambdabar}) and \eqref{equ:mubar}.
\STATE \textbf{\textit{Correction Step:}}
\STATE CSOs, SESOs, and DSO further update the energy profile $(p^{k+1}_{d,i,t},{p}^{k+1}_{b,i,t},{p}^{k+1}_{g,i,t}, p^{k+1}_{b,g,t}, p^{k+1}_{g,b,t})$, and the prices $\lambda_{i,t}^{k+1},\forall i,\forall t$, $\mu_{b,t}^{k+1},\forall b,\forall t$ according to (\ref{equ:pditk+1})-(\ref{equ:mubtk+1}), and then broadcast them to each other.
\STATE {Set $k=k+1$}
\UNTIL {convergence stopping criterion (\ref{equ:r}) is satisfied.}
\end{algorithmic}
\end{algorithm}


\emph{Condition A1}: The parameters $\tau$ and $\alpha$ satisfy that
\begin{align}
    M=\left(\begin{array}{ccc}
        2-2\alpha-\alpha\tau  & 1-\alpha-\alpha \tau  & -1+\alpha \\
        1-\alpha-\alpha \tau  & 2-2\alpha & -1+\alpha \\
        -1+\alpha & -1+\alpha & {2-\alpha} \\
    \end{array}\right) \succ 0. \nonumber
\end{align}

\begin{theorem} \label{thm-1}
When Condition A1 holds, Algorithm 1 converges to the system equilibrium, which is also the optimal solution of \eqref{eq:central}.
\end{theorem}

The proof of Theorem \ref{thm-1} can be found in Appendix \ref{appendix-B}. In fact, condition A1 is easy to meet. For example, we can let $\tau=0$ and $\alpha \in (0,1)$.
Compared with the traditional ADMM, the proposed distributed coordination mechanism has a convergence guarantee. The performance of the proposed mechanism will be further tested in the following simulations.

\emph{Remark:} Multi-block distributed optimization (including three-block problems) is an important area that has captured great attention in recent years. Though some researchers in the field of mathematics have proposed algorithms based on Gaussian feedback \cite{he2012alternating} and splitting method \cite{he2015splitting}, applying them to power system problems is not trivial. This is because those algorithms have strict restrictions on the parameter selection in order to ensure convergence, which requires an in-depth analysis of the problem structure and is highly case-by-case. To the best of our knowledge, there is no existing research applying the three-block algorithms in power system problems. Focusing on the coordination problem of multiple charging stations with shared energy storage in a distribution network, this paper gives the convergence condition theoretically in Theorem \ref{thm-1} and Appendix \ref{appendix-B}.

\section{Simulation Results and Discussions}\label{sec:result}
\begin{figure}[!htbp]
  \centering
  \includegraphics[width=0.4\textwidth]{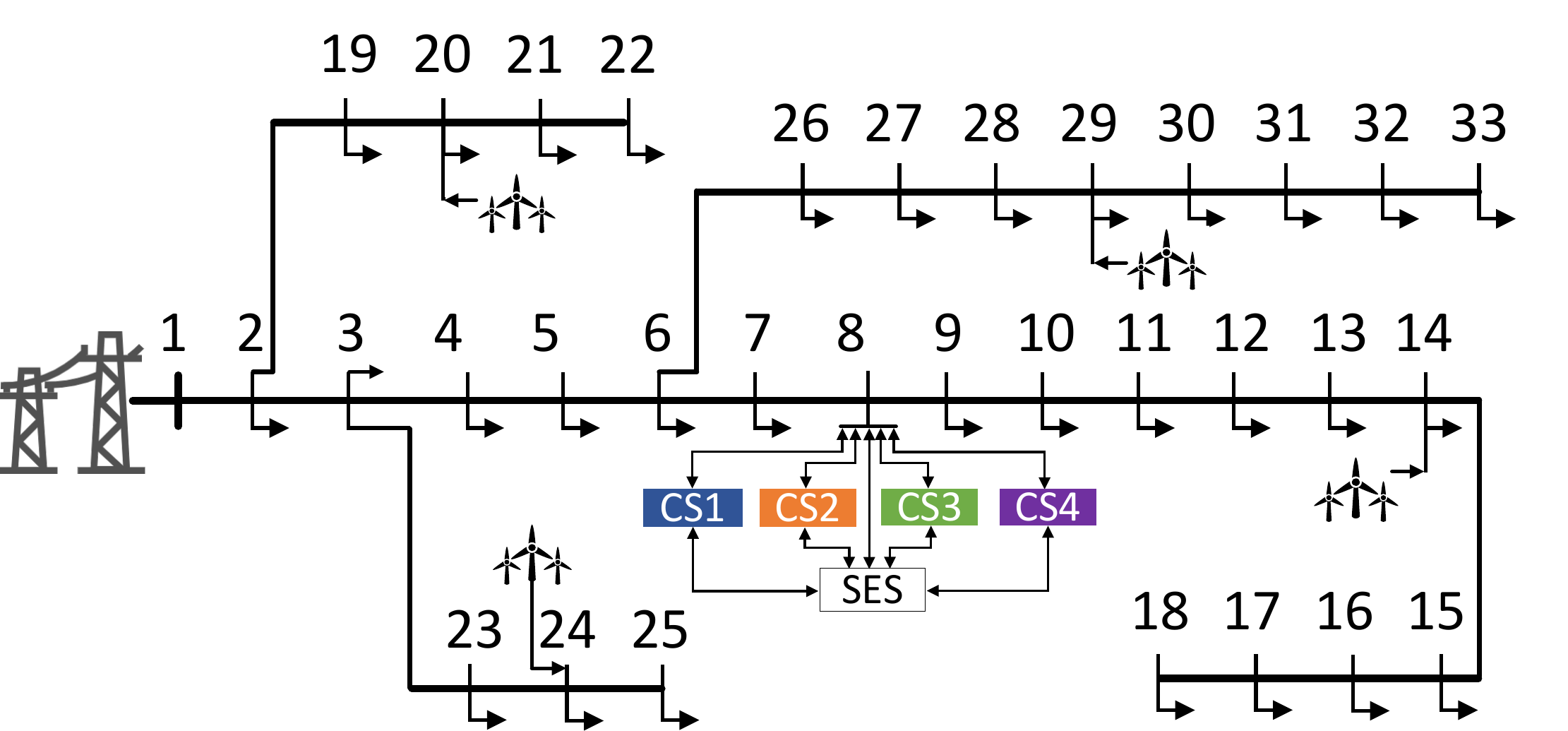}\\
  \caption{Modified IEEE 33-bus test system.}\label{fig:ieee33}
\end{figure}

In this section, we first evaluate the performance of the proposed distributed mechanism in an IEEE 33-bus test system \cite{ieee33}, which interconnects with four adjacent charging stations: CS1, CS2, CS3, and CS4, and the CSs share a common SES as shown in Fig. \ref{fig:ieee33}.
Then, the scalability of the proposed algorithm will be tested later using a system with more CSs and more SESs.

\subsection{System Setup}
All four charging stations are equipped with PV generations but in different capacities, and they are connected to a common shared energy storage. The relevant shared energy storage parameters are: the energy capacity of shared energy storage is $E_{b,cap}=650\text{ kWh}$ when it is fully charged, and its minimum/maximum energy is $E_{b}^{min}/E_{b}^{max}=0.1/0.9E_{b,cap}$, maximal charging/discharging power $p^{c,max}_{b}/p^{d,max}_{b}=0.3E_{b,cap}$, charging/discharging efficiency $\eta_d/\eta_c=0.95$, degradation cost coefficient $c_{b}=0.01$ \cite{wang2018incentivizing}. The inconvenience coefficient for charging station $c_{cs,i}$ takes a smaller value of 0.0001 to promote the utilization of EV charging flexibility \cite{cui2021community}.
Bus 1's power purchasing price from the utility grid $\lambda^b_{g,t}$ is set as the hourly electricity price from the PJM electricity market~\cite{pjm}. 
The price of electricity sold to the utility grid $\lambda^s_{g,t}$ is set to a constant value 0.01 \$/kWh.
For the distribution network, we set the upper and lower voltage magnitude limits at each bus as 1.06 p.u. and 0.94 p.u., respectively.

\begin{figure}[!htbp]
  \centering
  \includegraphics[width=0.4\textwidth]{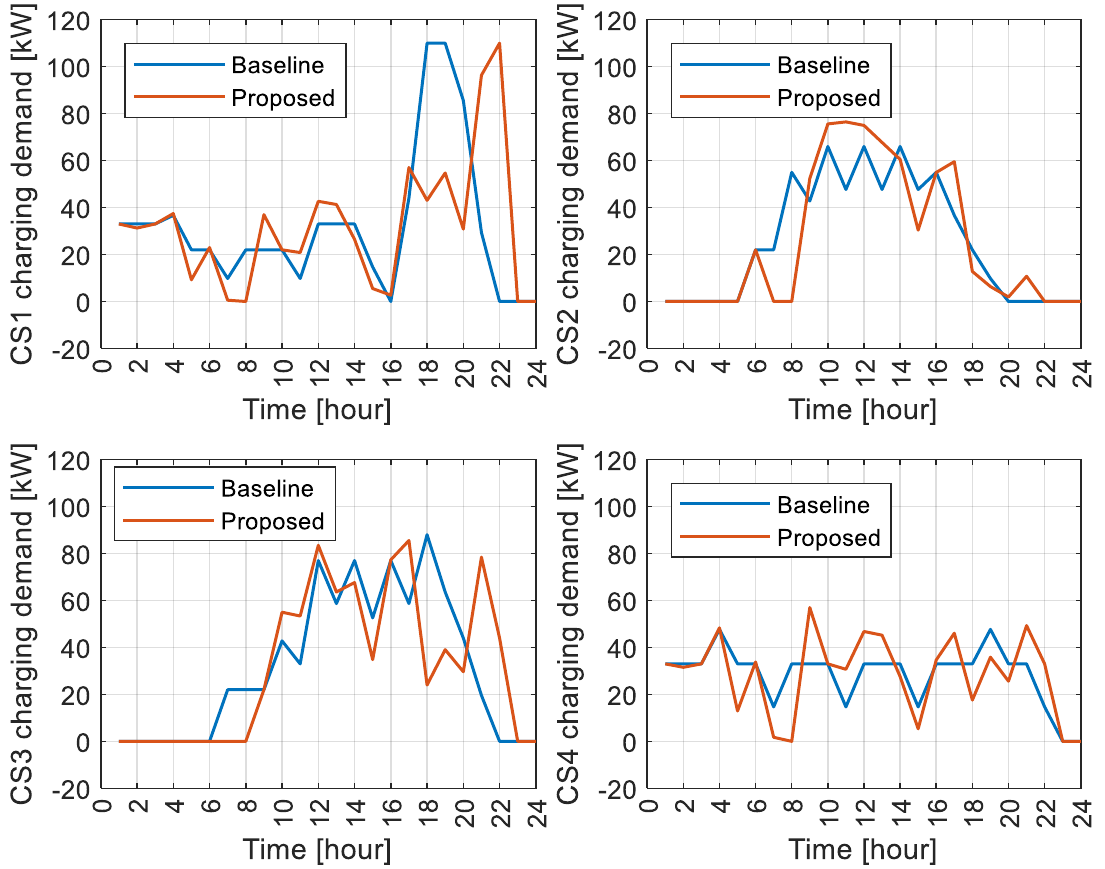}\\
  \caption{Charging demand at each charging station: baseline (CASAP way) vs. proposed.}\label{fig:demand}
\end{figure}

We consider four charging stations (CSs) located in different areas such as a residential area, a workplace area, a public leisure area, etc. \cite{gong2017load}. Thus, the charging demand pattern will show different characteristics, i.e., different charging load distributions over time. For example, the charging station in a residential area has more charging demands at night than during the day (e.g., CS1 in Fig. \ref{fig:demand}), while a workplace charging station exhibits the opposite load characteristics (e.g., CS2 in Fig. \ref{fig:demand}). The charging load of a charging station in a public leisure area also differs from the other two areas (e.g., CS3 and CS4 in Fig. \ref{fig:demand}). The number of EVs in CS1, CS2, CS3, and CS4 are 20, 16, 20, and 18, respectively, for a total of 74 EVs. We adopt the Tesla Model 3 EV with a 60 kWh battery capacity.

Fig. \ref{fig:demand} compares the total charging demand profiles under the baseline (see Trajectory 1 in Fig. \ref{fig:evcharging}, which is calculated by \eqref{equ:evbaseline}) and the proposed method (utilizes EV charging flexibility according to \eqref{eq:EV}). As we can see, the proposed method changes the baseline charging demand profiles, i.e., providing power flexibility to better meet the needs of the charging station and distribution network. For instance, the demand peak in CS1 is cut off and shifted to a later period when electricity prices are lower; The charging demand of CS2 is scheduled to increase at noon to consume sufficient PV power generation. To quantify the benefits of flexible EV charging, we will compare its impact on the total cost later.

\subsection{Effectiveness of the Distributed Coordination Algorithm}
\begin{figure}[!htbp]
  \centering
    \includegraphics[width=0.4\textwidth]{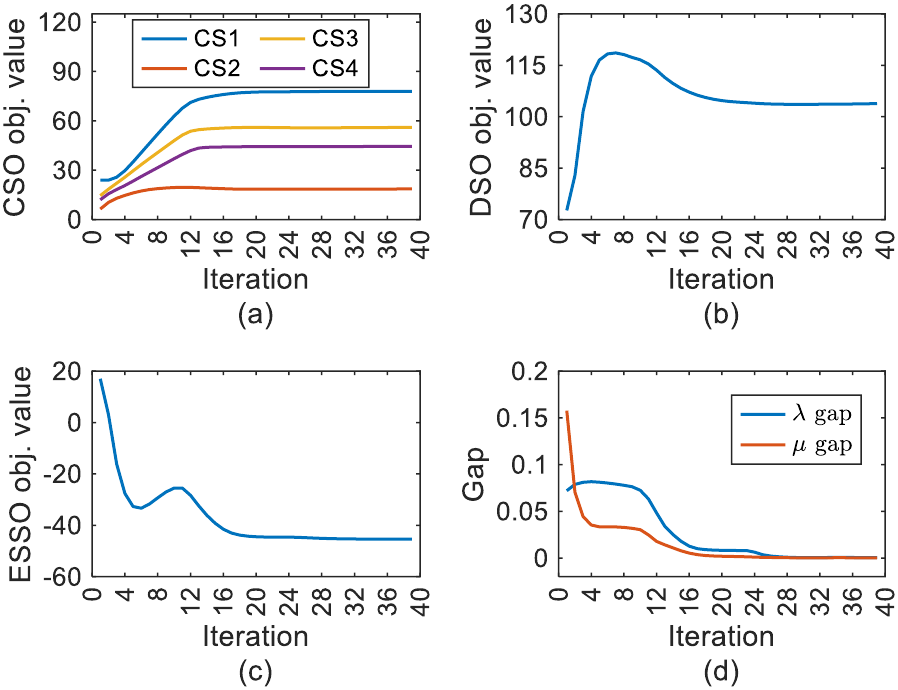} \\
  \caption{Convergence illustration. (a) CSO' objective function values. (b) DSO's objective function value. (c) SESO's objective function value. (d) Gaps.}\label{fig:converge}
\end{figure}

Fig. \ref{fig:converge} shows the convergence of the proposed distributed coordination mechanism. We illustrate the iterative process of CSOs', DSO's, and SESOs' objective function values, respectively, and also the error between two successive iterations. It can be found that the objective function values change rapidly in the first few iterations and then gradually converge. Overall, the algorithm only takes about 39 iterations to converge, showing a fast and acceptable rate of convergence.

\begin{figure}[!htbp]
  \centering
  \includegraphics[width=0.4\textwidth]{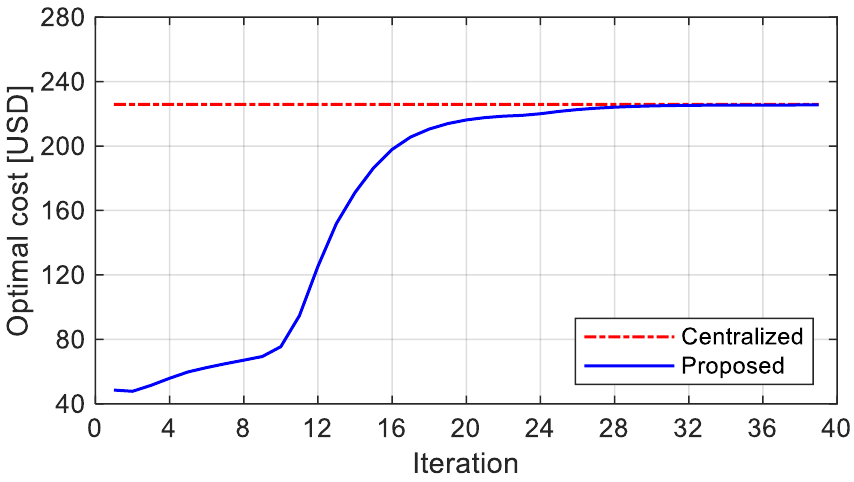} \\
  \caption{The optimal total cost convergence.}\label{fig:objiter}
\end{figure}

Theorem \ref{thm-1} claims that the proposed distributed coordination algorithm can converge to the centralized optimum. Here, we investigate this claim's effectiveness by comparing the proposed algorithm's obtained optimal cost and the centralized optimal cost, as shown in Fig. \ref{fig:objiter}. As the iteration proceeds, the optimal cost of the proposed algorithm gradually approaches the centralized optimal cost. This validates Theorem \ref{thm-1}.

\begin{figure}[!htbp]
  \centering
  \includegraphics[width=0.4\textwidth]{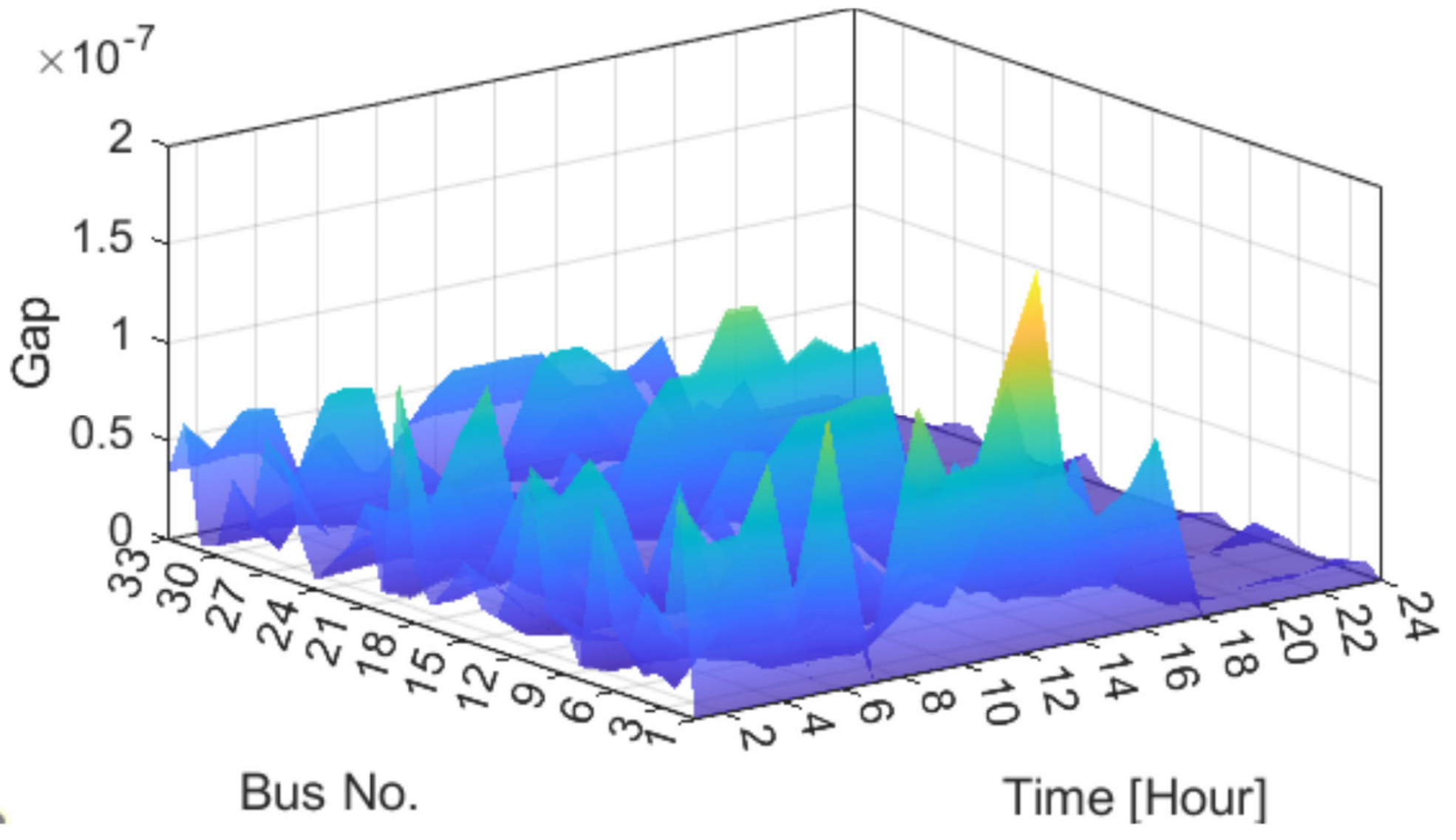}\\
  \caption{The gap at each bus under different periods.}\label{fig:vgap}
\end{figure}

At the equilibrium point of our coordinated mechanism, we check the exactness of the second-order cone relaxation. The differences between the left hand side $\ell_{n,j,t}$ and the right hand side $\frac{P_{nj,t}^2+Q_{nj,t}^2}{v_n(t)}$ of the inequality (\ref{equ:socp}) are illustrated in Fig. \ref{fig:vgap}. All gap values are less than $1\times10^{-6}$. Therefore, equality almost holds for the inequality constraint (\ref{equ:socp}), and the second-order conic relaxation is exact.

\subsection{Performance Evaluation}
Three widely used baselines in the literature are compared to show the advantage of the proposed distributed coordination algorithm.
\begin{itemize}
  \item The first baseline (B1) refers to the case with no individual nor shared energy storage in charging stations. This baseline is basically used to demonstrate the significance of deploying energy storage in charging stations. Its optimization problem is formulated as:
    \begin{equation}\label{equ:objB1}
        \begin{aligned}
            \min~&\sum_{i\in\mathcal{I}}C_{cs,i}+C_{ds}\\
            \text{s.t.}~& p_{d,i,t}+p_{g,i,t}= p_{pv,i,t},\forall i,\forall t\\
            ~ &  (\ref{equ:pc})-(\ref{equ:Evtd}),(\ref{equ:pinpf})-(\ref{equ:pflvub}).
        \end{aligned}
    \end{equation}
  \item The second baseline (B2) adopts individual energy storage in each charging station instead of sharing a common energy storage. Compared with B1, this baseline adds energy storage for each charging station. However, compared with the proposed shared-energy-storage architecture, this widely used individual energy storage architecture may suffer from a high investment cost and lower utilization rate of batteries. In B2, we divide the shared energy storage's capacity equal to the four charging stations. The optimization problem of B2 is formulated as follows:
    \begin{subequations}\label{equ:objB2}
        \begin{gather}
            \min~\sum_{i\in\mathcal{I}}C_{csb,i}+C_{ds},
        \end{gather}
        where
        \begin{gather}
        C_{csb,i}=C_{cs,i}+C_{bat,i},\\
        C_{bat,i}=c_{b}\sum\limits_{t\in\mathcal{T}}\left(p^d_{b,i,t}+p^c_{b,i,t}\right)\Delta t,
        \end{gather}
        subject to
        \begin{gather}
           p_{d,i,t}+p_{b,i,t}+p_{g,i,t}=p_{pv,i,t},\forall i,\forall t\\
            (\ref{equ:pc})-(\ref{equ:Evtd}),\\
            \text{constraints for the four energy storages,}\\
            (\ref{equ:pinpf})-(\ref{equ:pflvub}).
        \end{gather}
    \end{subequations}
    \item The third baseline (B3) adopts the proposed shared energy storage architecture but with an inelastic charging demand profile determined by \eqref{equ:evbaseline}, i.e., without utilizing EV power flexibility.
\end{itemize}

The proposed distributed coordination mechanism and the above three baselines are compared as follows.

\begin{table*}[!hbtp]
  \centering
  \caption{Cost comparison between baseline and proposed mechanism (Unit: USD).}\label{tab:compare}
    \begin{tabular}{cccccccccc}
    \toprule
          & \multicolumn{5}{c}{Charging station}  & Shared energy storage & Distribution network & \multirow{2}{*}{Total cost} & \multirow{2}{*}{Reduction} \\
          & \multicolumn{1}{c}{$C_{cso,1}$} & \multicolumn{1}{c}{$C_{cso,2}$} & \multicolumn{1}{c}{$C_{cso,3}$} & \multicolumn{1}{c}{$C_{cso,4}$} & \multicolumn{1}{c}{$\sum C_{cso,i}$} & $C_{seso,1}$ & $C_{dso}$ &       &  \\
    \midrule
    B1    & 74.72 & 10.48 & 53.81 & 40.22 & 179.23 & 0.00  & 108.80 & 288.04  & - \\
    B2    & 56.75 & 6.96  & 37.77 & 25.95 & 127.43 & 0.00  & 113.22 & 240.65  & 16.45\% \\
    B3    & 80.09 & 14.37 & 59.42 & 40.97 & 194.84 & -75.66 & 127.27 & 246.45  & 14.44\% \\
    Proposed & 69.29 & 10.43 & 49.88 & 37.40 & 167.00 & -45.38 & 103.84 & 225.46  & 21.72\% \\
    \bottomrule
    \end{tabular}%
\end{table*}%

\begin{table*}[htbp]
  \centering
  \caption{Profit allocation between CSO, SESO, and DSO under the proposed mechanism (Unit: USD)}  \label{tab:profit}%
    \begin{tabular}{cccccc|cccc|cccc}
    \toprule
    \multicolumn{6}{c|}{Charging station}        & \multicolumn{4}{c|}{Shared energy storage} & \multicolumn{4}{c}{Distribution network} \\
          & $C_{cs,i}$ & +$C_{i\text{-}b}$ & +$C_{i\text{-}g}$ & =$C_{cso,i}$ & $\sum C_{cso,i}$ & $C_{bat,1}$ & +$C_{b\text{-}i}$ & +$C_{b\text{-}g}$ & =$C_{seso,1}$ & $C_{ds}$ & +$C_{g\text{-}i}$ & +$C_{g\text{-}b}$ & =$C_{dso}$ \\
    \midrule
    CS1   & 7.94  & 0.48  & 60.87 & 69.29 & \multirow{4}[2]{*}{167.00} & \multirow{4}[2]{*}{14.24} & -0.48 & \multirow{4}[2]{*}{-77.48} & \multirow{4}[2]{*}{-45.38} & \multirow{4}[2]{*}{182.24} & -60.87 & \multirow{4}[2]{*}{77.48} & \multirow{4}[2]{*}{103.84} \\
    CS2   & 6.22  & -6.48 & 10.68 & 10.43 &       &       & 6.48  &       &       &       & -10.68 &       &  \\
    CS3   & 7.84  & -4.95 & 47.00 & 49.88 &       &       & 4.95  &       &       &       & -47.00 &       &  \\
    CS4   & 6.98  & -6.91 & 37.34 & 37.40 &       &       & 6.91  &       &       &       & -37.34 &       &  \\
    \bottomrule
    \end{tabular}%
\end{table*}%

\subsubsection{Cost Comparison}
TABLE \ref{tab:compare} summarizes the total cost and each cost item under the four different mechanisms.
First, we compare the total cost\footnote{The total cost stands for ``The total cost of the whole system running over a period of time'', i.e., from $t=1$ to $t=T$. Specifically, it is the sum of the costs of CSO, SESO, and DSO, i.e., $\sum_{i\in I} C_{cso,i}+\sum_{b\in\mathcal{B}} C_{seso,b}+C_{dso}$.}. With the simplest setting (no individual or shared energy storage and only flexible EV charging enabled), B1 has the highest total cost. Thanks to equipping the charging station with individual energy storage, the baseline B2 significantly decreases the total cost by 16.45\%. This indicates the importance of deploying energy storage at charging stations.
However, B2 adopts an individual energy storage architecture, which is hard to make full use of the complementary feature between different charging stations and may result in a low energy storage utilization rate.
In contrast, with a shared energy storage architecture, the proposed method achieves the lowest total cost with a cost reduction of 21.72\%.
Hence, the proposed shared energy storage architecture outperforms the individual energy storage architecture B2. That is partly because it can fully exploit the energy storage resources, as found in the next subsection.
In addition, the operation cost of the distribution network under Baseline B1 is 108.8 USD, which is 5\% higher than that under the proposed model (103.84 USD). This demonstrates that the distribution network can benefit from including energy storage.
We also discuss the impact of EV flexibility on the total operation cost.
B3 has the same shared energy storage architecture as the proposed method but without flexible EV charging. As a result, B3 leads to a higher total cost than the proposed one, demonstrating the significance of flexible EV energy resources. Nevertheless, B3 still benefits the overall system with a considerable total cost reduction compared with B1.
Moreover, the profit of SESO under B3 is more than that under the proposed method (75.66 USD $>$ 45.38 USD). As B3 lacks EV power flexibility, shared energy storage is the main flexible resource in the system. Hence, shared energy storage is used more frequently and provides more services to the CSOs and the DSO to reduce their operational costs. That is why the SESO obtains more profits.

\subsubsection{Profit Allocation}
TABLE \ref{tab:profit} further shows the detailed profit allocation among the charging stations, shared energy storage, and distribution system under the proposed mechanism. Here, $C_{x\text{-}y}$ represents the trading cost agent $x$ pays to agent $y$; $i$ is the index of charging stations, $b$ is the index of shared energy storage, and $g$ refers to the distribution power grid. A negative $C_{x-y}$ means agent $x$ is actually earning profit from agent $y$. Moreover, we have $C_{x\text{-}y}+C_{y\text{-}x}=0,\forall x,\forall y$. This is because the payment from agent $x$ to agent $y$ is the profit agent $y$ gets from agent $x$.

We first analyze the profit allocation between each CS and the SES. $C_{i\text{-}b}$ is positive for CS1, while for CS2-CS4, $C_{i\text{-}b}$s are negative. This means that CS1 is paying for the shared energy storage while CS2-CS4 are gaining profits. Besides, $\sum_{i=1}^4 C_{i\text{-}b}=-17.86$ USD is negative, showing that the charging station system (consisting of CS1-CS4) obtains profits from trading with the SES. For the profit allocation between the CSs and the grid, all CSs pay the grid for buying electricity. For the profit allocation between the SES and the grid, $C_{b\text{-}g}$ is negative, i.e., the SES obtains profits by trading with the grid. Moreover, the total operation cost for the SES is -45.38 USD, indicating that the SES can benefit from participating in this trading.

\subsubsection{Internal Power Distribution of Each Charging Station}
\begin{figure}[!htbp]
  \centering
    \includegraphics[width=0.49\textwidth]{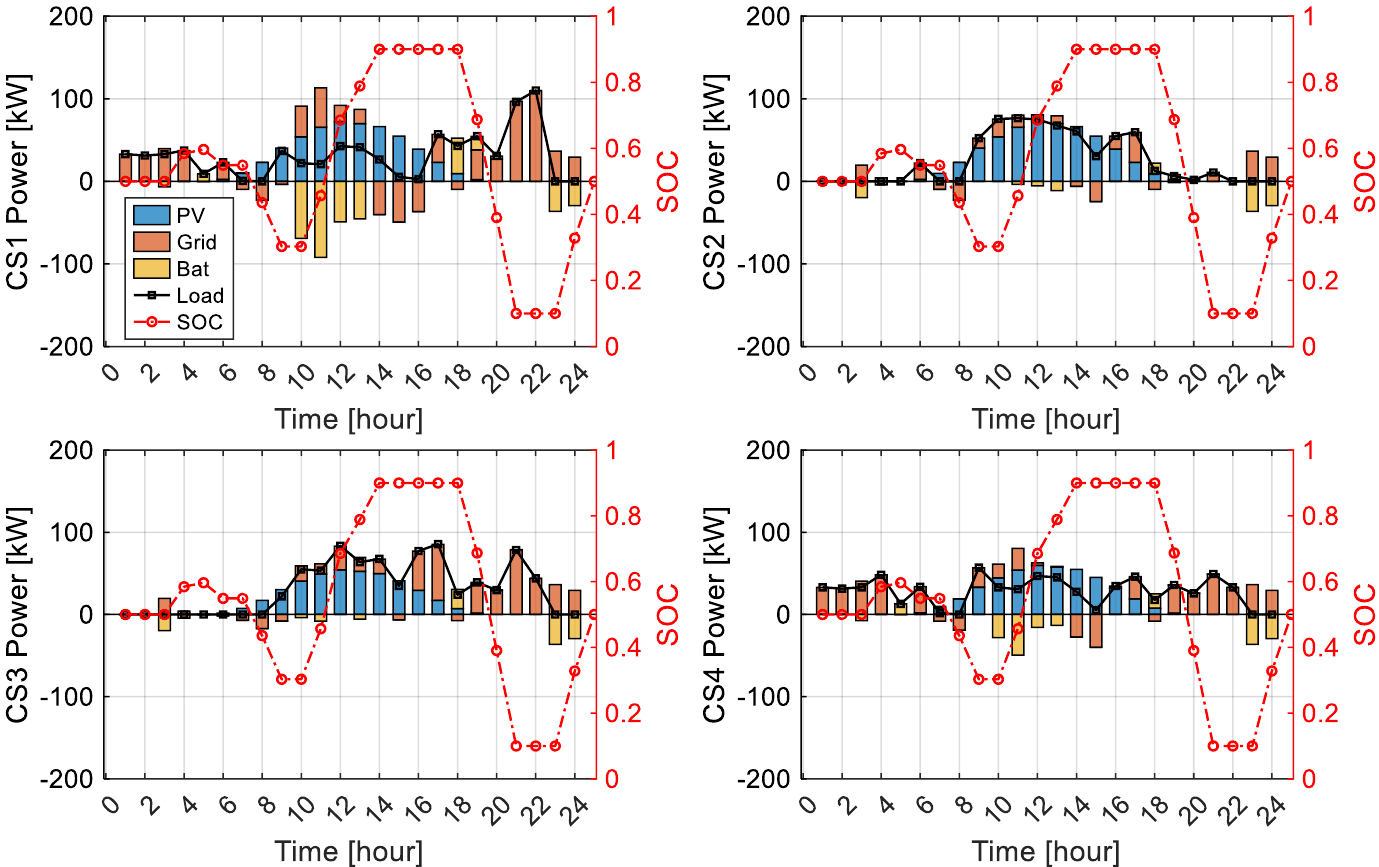} \\
  \caption{Proposed shared energy storage architecture: Power distribution inside each charging station.}\label{fig:power}
\end{figure}

\begin{figure}[!htbp]
  \centering
  \includegraphics[width=0.49\textwidth]{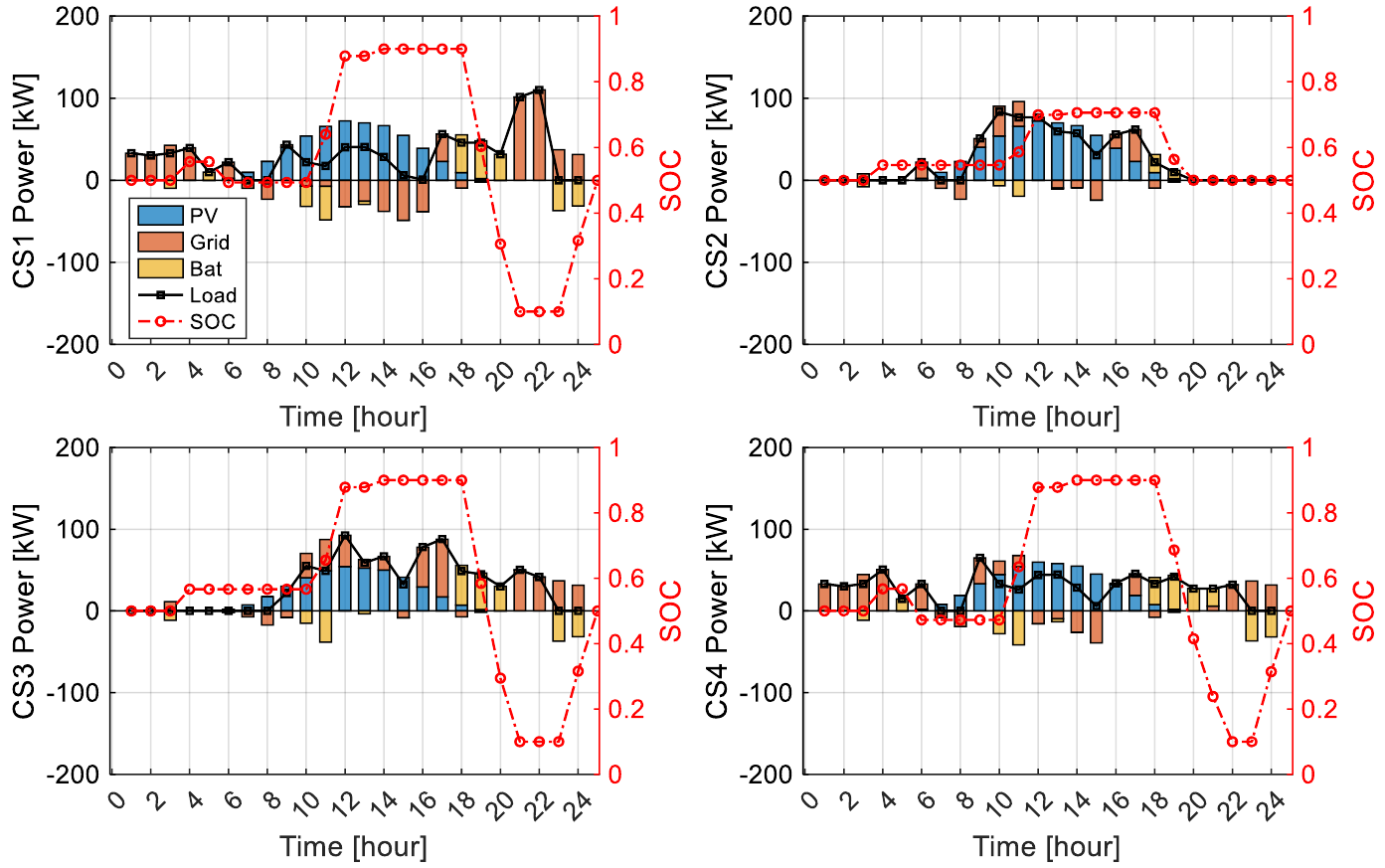} \\
  \caption{B2 individual energy storage architecture: Power distribution inside the four charging stations.}\label{fig:basepower}
\end{figure}

We investigate the internal power distribution of each charging station to identify the impact of shared energy storage.
Fig. \ref{fig:power} shows the power distribution among PV, grid, and shared energy storage within each charging station under the proposed method, as well as the time-varying SOC of the shared energy storage.
For comparison, Fig. \ref{fig:basepower} shows the power distribution under the baseline B2, i.e., the individual energy storage architecture.
Looking at the energy storage's SOC level, one major difference can be found: the energy storage of CS2 is underutilized in B2; while under the proposed shared energy storage architecture, this can be avoided due to the shared energy storage simultaneously serving multiple charging stations with diverse charging demand, and it experiences a complete charging and discharging cycle.
Therefore, it can be inferred that the utilization rate of energy storage can be greatly promoted through the shared energy storage architecture.

\subsubsection{Impact of energy storage Capacity}
\begin{figure}[!htbp]
  \centering
  \includegraphics[width=0.4\textwidth]{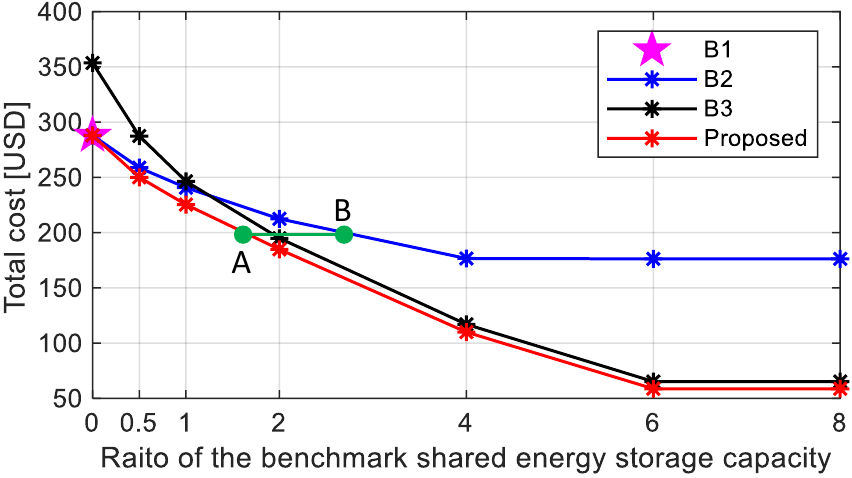} \\
  \caption{Impact of shared energy storage capacity on the total cost.}\label{fig:batcap}
\end{figure}

We further investigate the impact of shared energy storage capacity on the total cost of the overall system.
Fig. \ref{fig:batcap} shows the total cost trend as the (shared) energy storage capacity increases under different settings.
We first analyze the proposed case.
The trend curve presents a nonlinear relationship. When the energy storage capacity is zero, the proposed shared energy storage architecture degenerates to the baseline B1, resulting in a very high total cost at this point. As the energy storage capacity increases, the total cost drops rapidly and remains unchanged. It means that very large shared energy storage is unnecessary.
Considering that the investment cost of energy storage is still high, this trend curve is helpful in providing guidance on the sizing of shared energy storage.
For B2, the individual energy storage architecture has a similar trend curve, but its total cost is always larger than the proposed one, which validates the advantage of shared energy storage architecture again. The cost difference between B2 and the proposed model can be viewed as the value of adopting shared energy storage compared to individual energy storage.
Note that we follow the rule that the sum of the capacity of individual energy storage in each charging station is equal to the shared energy storage capacity. The individual energy storage capacity is set as the shared energy storage capacity divided by four. Therefore, as the shared energy storage capacity increases, the individual energy storage capacity also increases. Different energy storage capacity leads to different total cost, rather than a fixed point like B1, as shown in Fig. \ref{fig:batcap}.
For B3, it has the highest total cost when energy storage capacity is less than $1.2E_{b,cap}$ due to the lack of EV flexibility. Moreover, as the SES capacity increases, the total cost difference between B3 and the proposed model gets smaller. Similarly, this cost gap implies the value of EV flexibility, which is influenced by shared energy storage capacity. For example, at the point of $0.5E_{b,cap}$, where the shared energy storage capacity is small, leveraging EV flexibility can significantly reduce the total cost by approximately 40 USD. While at $2E_{b,cap}$, where energy storage capacity becomes larger, the cost only decreases by approximately 10 USD.
Overall, the proposed method can always achieve the best result.

Furthermore, the advantage of the proposed system architecture with shared energy storage can be justified by showing the potential investment cost reduction. As shown in Fig. \ref{fig:batcap}, with the same total energy storage capacity, the total cost incurred by the proposed architecture is lower than B2 (individual energy storage architecture). Or in another way: to achieve the same total cost, the proposed architecture (point A) requires a smaller total energy storage capacity than the individual energy storage architecture (point B). Therefore, the reduced energy storage capacity can be considered a decrease in investment costs.

Additionally, the shared energy storage capacity needed is also influenced by the generation and load patterns of the connected charging stations. To investigate this impact, we test the performance of the proposed model and Benchmark B2 using a case with four identical charging stations (the same PV power generation and EV charging pattern). As we can imagine, the diminished net load complementarity of charging stations will lead to a higher energy storage capacity requirement. We find that under the same total cost, the shared energy storage can save 9\% capacity compared to the individual energy storage. Even if there is no complementarity in the external conditions, charging stations can complement each other through flexible charging, making the capacity required for shared energy storage smaller. This also implies that diverse and complimentary charging loads are desired for shared energy storage.

\subsubsection{Impact of constraint \eqref{equ:soct1tT}}
It is worth noting that, though we include the restriction \eqref{equ:soct1tT} in our model to be more practical, the proposed method can still be applied, and the proposition and theorem still hold when removing it. To show this, we give the simulation results using a model without this restriction \eqref{equ:soct1tT} in Table \ref{tab:batE0}. The total cost becomes smaller without the constraint \eqref{equ:soct1tT}. This is because that without \eqref{equ:soct1tT}, at the end of the dispatch cycle, the shared energy storage can be discharged to an energy level lower than the initial level $E_0$ to earn more profits. However, this is at the expense of paying more in the next dispatch cycle to recharge to $E_0$.

\begin{table}[htbp]
  \centering
  \caption{Total cost comparison with/without constraint (15) (Unit: USD)}  \label{tab:batE0}%
    \begin{tabular}{cccc}
    \toprule
    & Shared energy                & Shared energy               & Individual energy \\
    & storage with \eqref{equ:soct1tT} & storage w/o \eqref{equ:soct1tT} & storage w/o \eqref{equ:soct1tT}\\
    \midrule
    Total cost   & 225.46  & 184.01 & 209.58\\
    \bottomrule
    \end{tabular}%
\end{table}%

\subsection{Impact of Coefficient Value}
\begin{figure}[!htbp]
  \centering
  \includegraphics[width=0.35\textwidth]{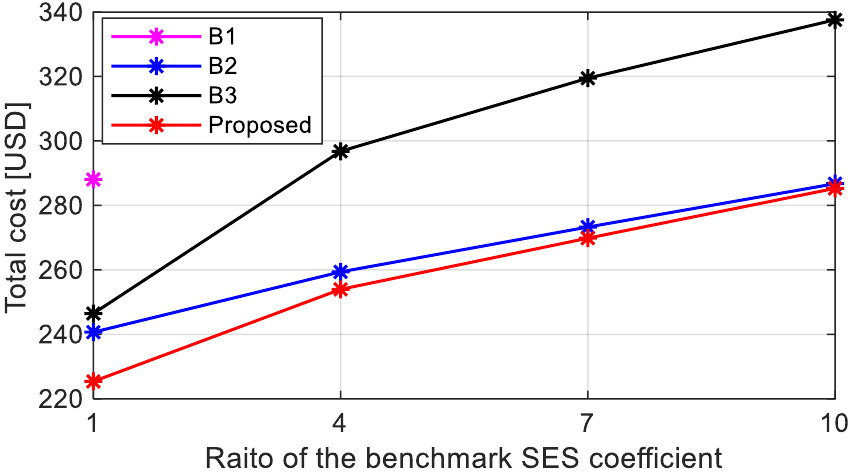} \\
  \caption{The impact of SES coefficient value on the total cost.}\label{fig:batcoef-total}
\end{figure}

\begin{figure}[!htbp]
  \centering
  \includegraphics[width=0.35\textwidth]{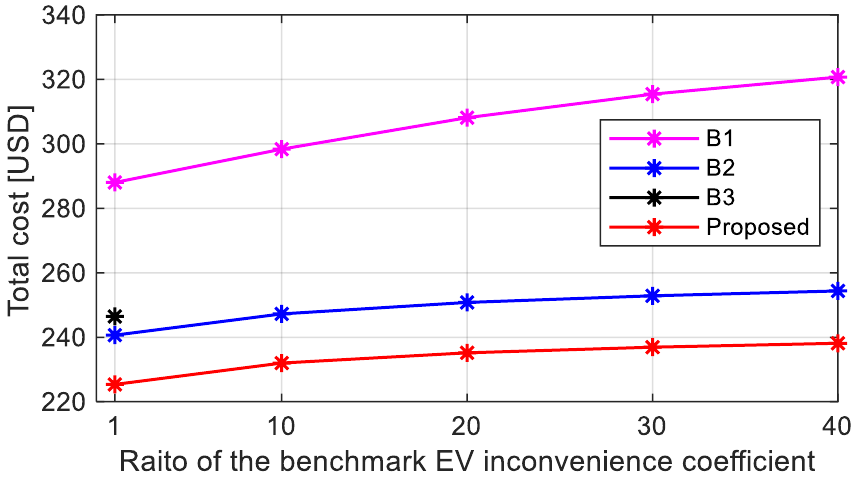} \\
  \caption{The impact of EV inconvenience coefficient value on the total cost.}\label{fig:evcoef-total}
\end{figure}

For energy storage, the cost coefficient $c_b$ is widely used to measure degradation. We used the typical parameter value given in \cite{wang2018incentivizing}. Likewise, for EV charging, the inconvenience coefficient $c_{cs,i}$ is used to measure the inconvenience caused by the deviation from the baseline CASAP trajectory, and we refer to \cite{cui2021community} for its typical value. Fig. \ref{fig:batcoef-total} and \ref{fig:evcoef-total} show the impact of the two cost coefficients on the total system cost. As the cost coefficient increases, the total cost of all methods increases. But our proposed method still outperforms other methods.

\begin{figure}[!htbp]
  \centering
  \includegraphics[width=0.35\textwidth]{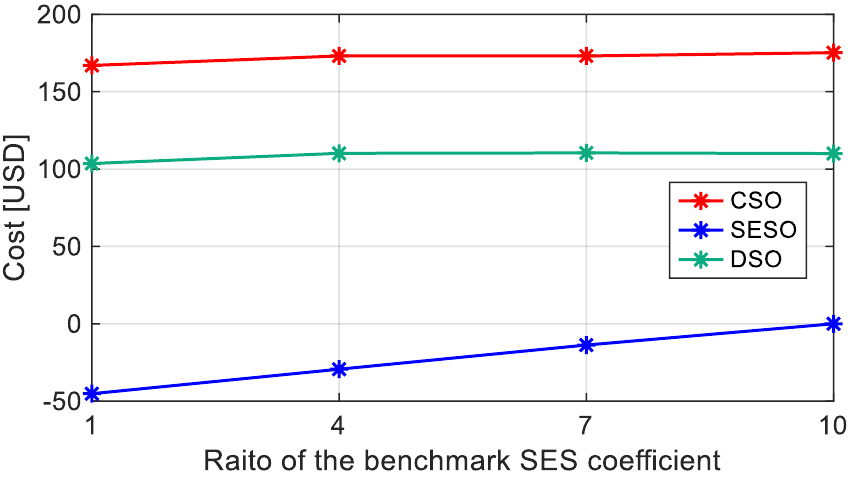} \\
  \caption{The impact of SES coefficient value on three costs.}\label{fig:batcoef-3}
\end{figure}

\begin{figure}[!htbp]
  \centering
  \includegraphics[width=0.35\textwidth]{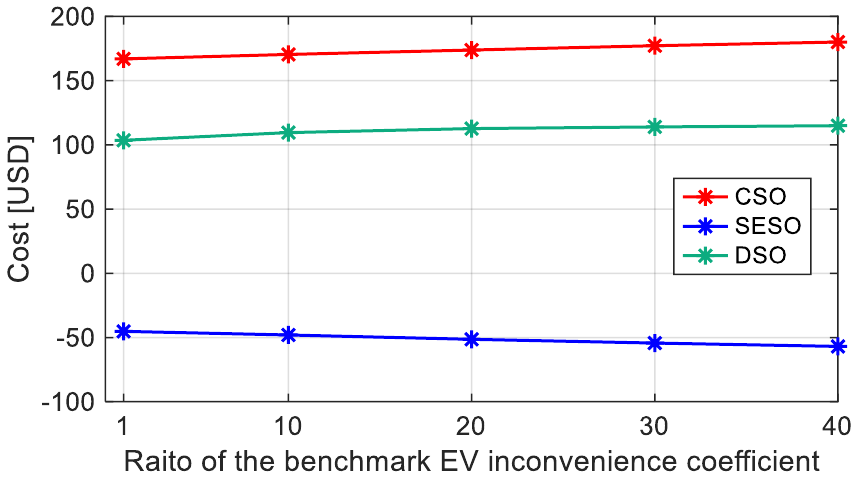} \\
  \caption{The impact of EV inconvenience coefficient value on three costs.}\label{fig:evcoef-3}
\end{figure}

At the same time, we also investigate how the costs of CSO, SESO, and DSO change with the increase of battery degradation coefficient under the proposed method. As seen in Fig. \ref{fig:batcoef-3}, SESO's cost increase rapidly, and CSO's and DSO's costs increase slightly. This indicates that the battery coefficient affects not only SESO cost but also CSO and DSO costs.
Similarly, Fig. \ref{fig:evcoef-3} shows the impact of EV inconvenience coefficient value on the CSO cost, SESO cost, and DSO cost. The costs of CSO and DSO increase while the cost of SESO decreases. That is because using EV charging flexibility becomes expensive and difficult, and the system shifts to using shared energy storage, resulting in more benefits for the SESO.

\subsection{Scalability}
\begin{figure}[!htbp]
  \centering
  \includegraphics[width=0.4\textwidth]{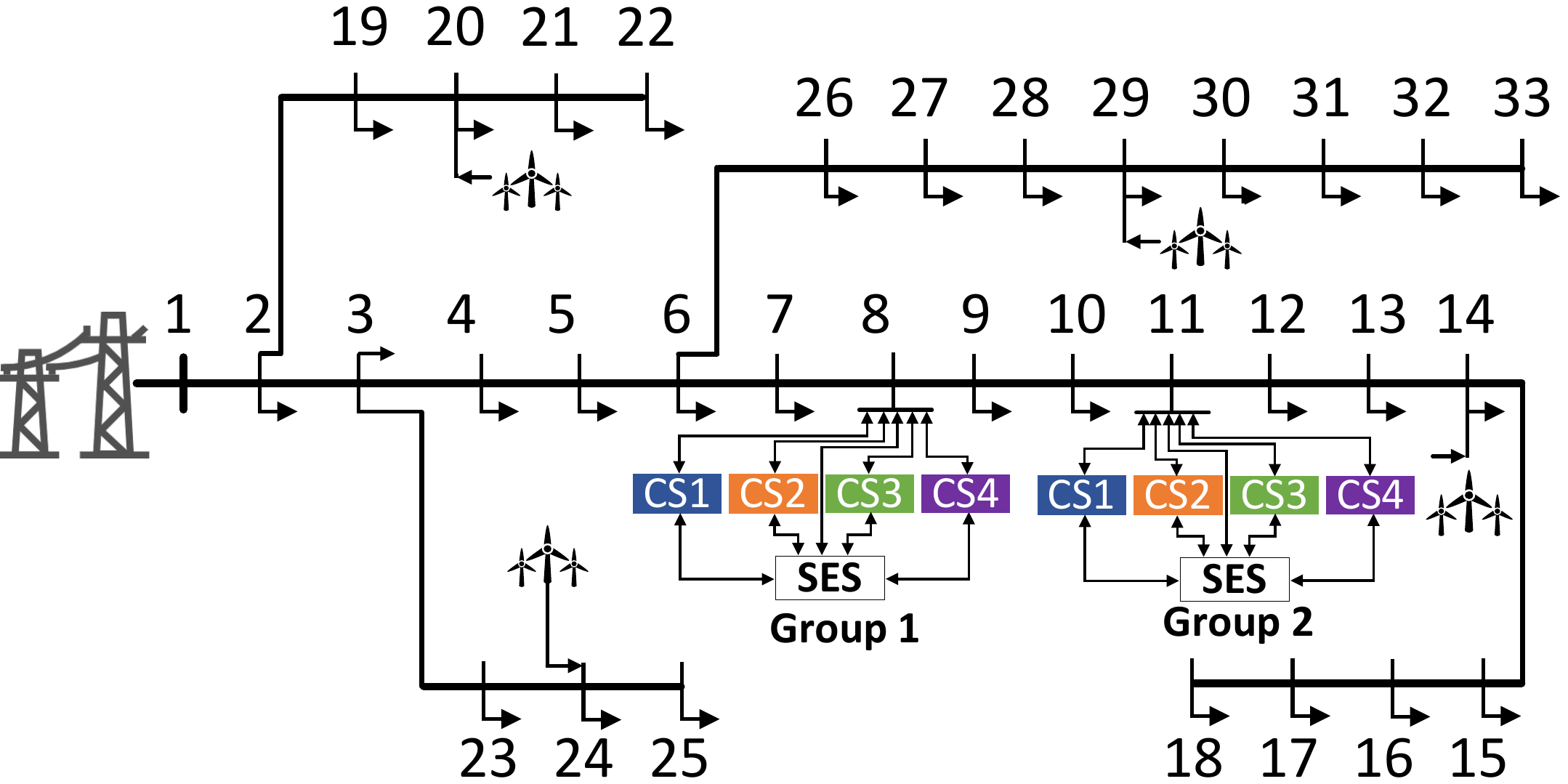} \\
  \caption{The case with two shared energy storages.}\label{fig:case2}
\end{figure}

\begin{figure}[!htbp]
  \centering
  \includegraphics[width=0.45\textwidth]{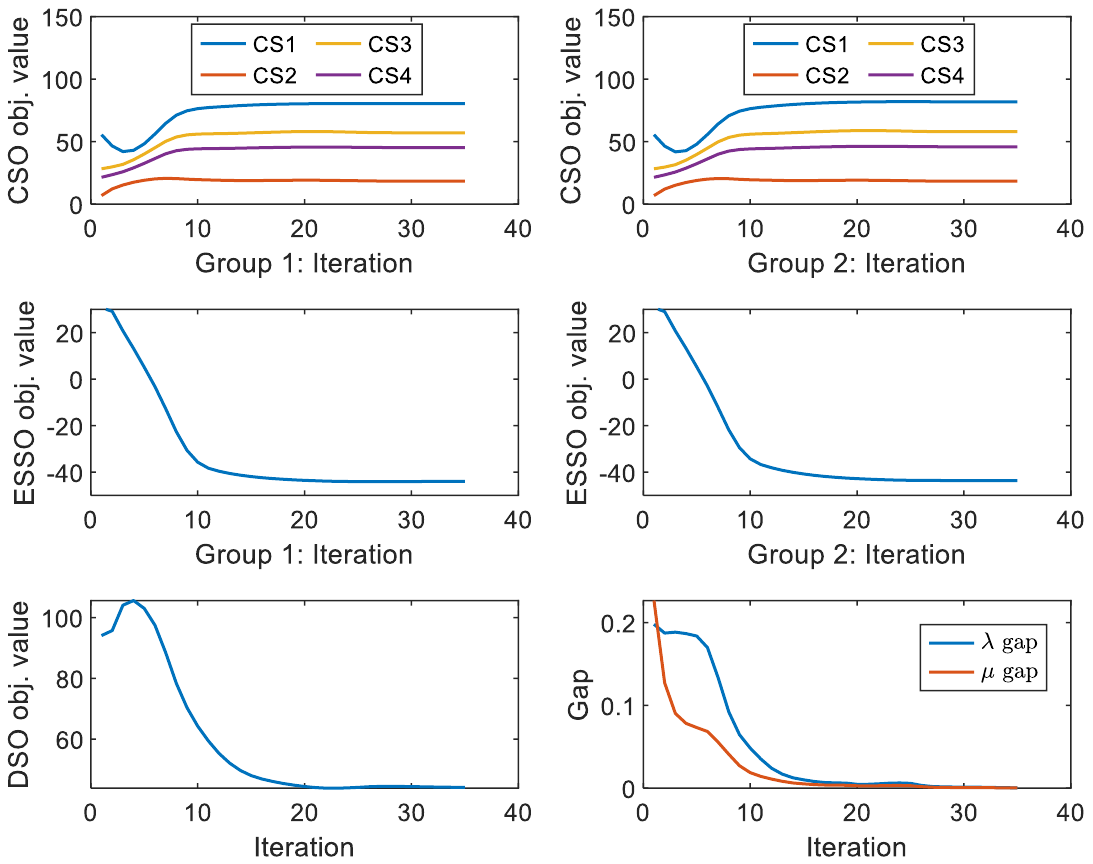} \\
  \caption{Iterations for the case with two shared energy storages.}\label{fig:case2iter}
\end{figure}

So far, we have investigated the effectiveness and performance of the proposed algorithm through a case of one shared energy storage and four charging stations.
Here, we further study the case that includes two shared energy storages (SESs), each connected to four charging stations in the distribution network, as shown in Fig. \ref{fig:case2}.
Fig. \ref{fig:case2iter} shows a good convergence performance for each objective value and the gap. This case verifies that the proposed algorithm can be extended to a case with multiple shared energy storages and charging stations.

\begin{figure}[!htbp]
  \centering
  \includegraphics[width=0.4\textwidth]{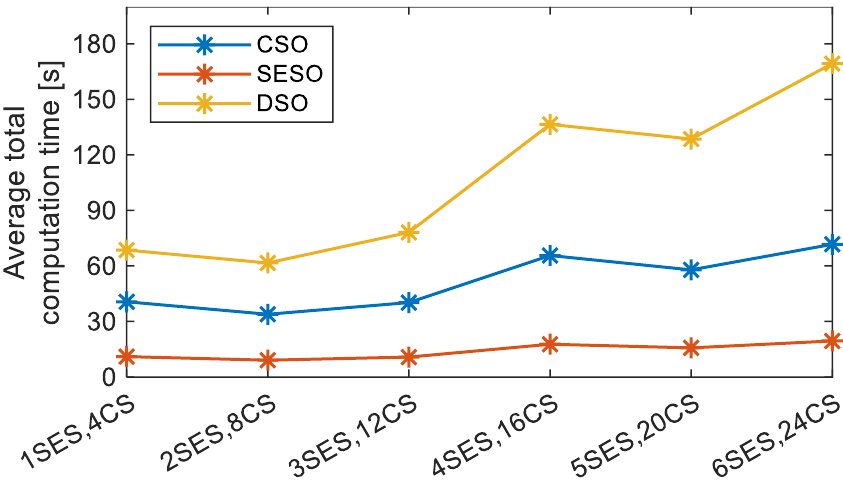} \\
  \caption{Scalability: Computational time vs. a number of shared energy storages and charging stations.}\label{fig:scala}
\end{figure}

To show the scalability of the proposed distributed algorithm, we compare the computational time under different settings (with the different numbers of groups of shared energy storage and charging stations): 1SES4CS, 2SES8CS, 3SES12CS, 4SES16CS, 5SES20CS, and 6SES24CS, where 6SES24CS means 6 shared energy storages, 24 charging stations (1776 EVs in total) in the IEEE 33-bus.
In fact, the proposed mechanism runs in a distributed manner, which means that different agents can do their local computations in parallel. Hence, the increase in scene complexity has little impact on each agent's average computational time. As we can see from Fig. \ref{fig:scala}, with more charging stations and shared energy storage, the computational time slightly increases but remains within an acceptable range. The DSO needs to take longer because of the growing number of charging stations, but the growth is acceptable. Moreover, the computational time for each agent is somehow uncertain and can be affected by various factors, e.g., the nodal to which charging stations are connected, the power supply/demand condition of the overall system, etc. It is possible that the computational time under (5SES, 20CS) is smaller than that under (4SES, 16CS).

\section{Conclusion}\label{sec:conclu}
This paper proposes a shared energy storage architecture for EV charging stations in a distribution network. An equilibrium model is developed to characterize their interactions. We prove that the equilibrium exists and coincides with the centralized optimum when the trading prices are set as the value of certain dual variables at the optimum. To achieve equilibrium, a novel distributed coordination mechanism is proposed with proof of convergence. Simulation results demonstrate the effectiveness and scalability of the proposed algorithm and report the following findings:

(1) Compared with the individual energy storage architecture, the shared energy storage-based one can reduce the total cost by 5.27\% and improve the utilization of energy storage.

(2) The shared energy storage capacity can affect the total operational cost, but a very large energy storage size is unnecessary.

(3) The proposed algorithm has good scalability with increasing penetration of charging stations and shared energy storage.

\appendices

\makeatletter
\@addtoreset{equation}{section}
\@addtoreset{theorem}{section}
\makeatother
\setcounter{definition}{0} 
\renewcommand{\theequation}{A.\arabic{equation}}
\renewcommand{\thetheorem}{A.\arabic{theorem}}
\renewcommand{\thedefinition}{A\arabic{definition}}

\section{Proof of Proposition \ref{prop-1}}
\label{appendix-A}
The CSO's problem is equivalent to
\begin{subequations}
\begin{align}
    \min ~ & C_{cs,i} - \sum_{t \in \mathcal{T}} \lambda_{g,i,t}p_{g,i,t} -\sum_{t \in \mathcal{T}} \lambda_{b,i,t} p_{b,i,t}, \\
    \mbox{s.t.}~ & p_{d,i,t}+p_{g,i,t}+p_{b,i,t}=p_{pv,i,t}: \eta_{i,t}, \\
    ~ & \{p_{d,i,t},\forall t\} \in \mathcal{D}_i.
\end{align}
\end{subequations}
Based on the variational inequality, if $(p_{g,i}^*, p_{b,i}^*, p_{d,i}^*)$ \footnote{For simplicity, we ignore the index $t$, and $p_{g,i}^*$ is a collection of $p_{g,i,t}^*$ for all $t$; others symbols without a $t$ are defined similarly.} is the optimal solution and $\eta_{i}^*$ the corresponding dual variable, then for all $(p_{g,i}, p_{b,i}, p_{d,i},\eta_{i}) \in \mathbb{R}^T \times \mathbb{R}^T \times \mathcal{D}_i \times \mathbb{R}^T$, we have
\begin{subequations}\label{eq:optimality1}
\begin{align}
    C_{cs,i}(p_{d,i})-C_{cs,i}(p_{d,i}^*) + \sum_{t \in \mathcal{T}}\eta_{i,t}^* (p_{d,i,t}-p_{d,i,t}^*) \ge 0, \label{apeq:cso-pdit}\\
    (-\lambda_{g,i,t} + \eta_{i,t}^*) (p_{g,i,t}-p_{g,i,t}^*) \ge 0,\forall t, \label{eq:optimality1-2}\\
     (-\lambda_{b,i,t} + \eta_{i,t}^*) (p_{b,i,t}-p_{b,i,t}^*) \ge 0,\forall t, \label{eq:optimality1-3}\\
     \sum_{t \in \mathcal{T}} (\eta_{i,t}-\eta_{i,t}^*)(p_{d,i,t}^*+p_{g,i,t}^*+p_{b,i,t}^*-p_{pv,i,t}) \ge 0.  \label{apeq:cso-bala}
\end{align}
\end{subequations}

The SESO's problem is equivalent to
\begin{subequations}
\begin{align}
    \min ~ & C_{bat,b} + \sum_{t\in \mathcal{T}} \sum_{i \in \mathcal{I}_b} \lambda_{b,i,t} p_{b,i,t} + \sum_{t \in \mathcal{T}} \lambda_{b,t} p_{b,g,t}, \\
    \mbox{s.t.}~ & \{p_{b,i,t}, \forall i\in \mathcal{I}_b,\forall t;p_{b,g,t}, \forall t\} \in \mathcal{F}_b.
\end{align}
\end{subequations}
Based on the variational inequality, if $(p_{b,i}^{'},\forall i; p_{b,g}^*)$ is the optimal solution, then for all $(p_{b,i},\forall i;  p_{b,g}) \in \mathcal{F}_b$, we have
\begin{align}\label{eq:optimality2}
   ~ & C_{bat,b}(p_{b,i},\forall i \in \mathcal{I}_b, p_{b,g})- C_{bat,b}(p_{b,i}^{'},\forall i \in \mathcal{I}_b, p_{b,g}^*) \nonumber\\
    ~ & + \sum_{t\in \mathcal{T}} \sum_{i \in \mathcal{I}_b} \lambda_{b,i,t} (p_{b,i,t}-p_{b,i,t}^{'}) \nonumber\\
    ~ & + \sum_{t \in \mathcal{T}} \lambda_{b,t} (p_{b,g,t}-p_{b,g,t}^*)\ge 0. 
\end{align}

The DSO's problem is equivalent to
\begin{subequations}
\begin{align}
    \min~ & C_{ds}+\sum_{t\in \mathcal{T}}\sum_{i \in \mathcal{I}} \lambda_{g,i,t} p_{g,i,t} - \sum_{t \in \mathcal{T}} \sum_{b \in \mathcal{B}} \lambda_{b,t} p_{b,g,t}, \\
    \mbox{s.t.}~ & p_{g,b,t}+p_{b,g,t}=0: \phi_{b,t}, \\
    ~ & \{p_{g,i,t}, \forall i,\forall t; p_{g,b,t},\forall b,\forall t\} \in \mathcal{P}.
\end{align}
\end{subequations}
Based on the variational inequality, if $( p_{b,g}^{'}, \forall b; p_{g,i}^{'}, \forall i;  p_{g,b}^*,\forall b)$ is the optimal solution and $\phi_{b}^*,\forall b$ the corresponding dual variable, then for all $(p_{b,g}, \forall b; p_{g,i},\forall i; p_{g,b}, \phi_{b},\forall b) \in \mathbb{R}^{B\times T}  \times \mathcal{P} \times \mathbb{R}^{B \times T}$, we have
\begin{subequations}\label{eq:optimality3}
\begin{align}
    &C_{ds}(p_{g,i}, p_{g,b}) + \sum_{t \in \mathcal{T}} \sum_{i \in \mathcal{I}} \lambda_{g,i,t} (p_{g,i,t}-p_{g,i,t}^{'})\nonumber\\
    &-C_{ds}(p_{g,i}^{'}, p_{g,b}^*)+\sum_{t \in \mathcal{T}} \sum_{b \in \mathcal{B}} \phi_{b,t}^* (p_{g,b,t}-p_{g,b,t}^*) \ge 0, \label{apeq:dso-pgit}\\
    &(-\lambda_{b,t}+\phi_{b,t}^*)(p_{b,g,t}-p_{b,g,t}^{'})\ge 0, \forall b,\forall t \label{eq:optimality3-3}\\
    &\sum_{t\in \mathcal{T}} \sum_{b \in \mathcal{B}} (\phi_{b,t}-\phi_{b,t}^*)(p_{g,b,t}^*+p_{b,g,t}^{'}) \ge 0. \label{apeq:seso-pgbt}
\end{align}
\end{subequations}

The optimality conditions \eqref{eq:optimality1}, \eqref{eq:optimality2}, \eqref{eq:optimality3}, together with \eqref{eq:supply-demand} constitute the condition for a system equilibrium (SE).

Similarly, suppose $(\bar p_{g,i,t},\forall i,\forall t; \bar p_{b,i,t},\forall i,\forall t; \bar p_{b,g,t},\forall b,\forall t)$ is the optimal solution of \eqref{eq:central}, then for all $(p_{d,i},\forall i; p_{g,i},\forall i;  p_{g,b},\forall b;  p_{b,i},\forall i;  p_{b,g},\forall b; \lambda_{i}; \mu_{b}) \in \prod_i \mathcal{D}_i \times \mathcal{P} \times \prod_b \mathcal{F}_b \times \mathbb{R}^{I \times T} \times \mathbb{R}^{B \times T}$, we have
\begin{subequations}\label{eq:optimality4}
\small
\begin{gather}
    C_{cs,i}(p_{d,i})-C_{cs,i}(\bar p_{d,i})
    + \sum_t \bar \lambda_{i,t} (p_{d,i,t}-\bar p_{d,i,t}) \ge 0,\forall i, \label{apeq:cent-pdit}\\
    C_{ds}(p_{g,i},p_{g,b})-C_{ds}(\bar p_{g,i},\bar p_{g,b}) + \sum_t \sum_i \bar \lambda_{i,t} (p_{g,i,t}-\bar p_{g,i,t}) \nonumber\\
    + \sum_t \sum_b \bar \mu_{b,t} (p_{g,b,t}-\bar p_{g,b,t}) \ge 0, \label{apeq:cent-pgit}\\
    C_{bat,b}(p_{b,i},\forall i \in \mathcal{I}_b; p_{b,g})- C_{bat,b}(p_{b,i},\forall i \in \mathcal{I}_b; \bar p_{b,g}) \nonumber\\
    + \sum_t \sum_{i \in \mathcal{I}_b} \bar \lambda_{i,t} (p_{b,i,t}-\bar p_{b,i,t}) \nonumber\\
    + \sum_t \bar \mu_{b,t} (p_{b,g,t}-\bar p_{b,g,t}) \ge 0,\forall b, \label{apeq:cent-pbit}\\
    \sum_t \sum_i (\lambda_{i,t}-\bar \lambda_{i,t}) (\bar p_{d,i,t}+\bar p_{g,i,t}+\bar p_{b,i,t}-p_{pv,i,t}) \ge 0, \label{apeq:cent-bala}\\
    \sum_t \sum_b (\mu_{b,t}-\bar \mu_{b,t}) (\bar p_{g,b,t}+\bar p_{b,g,t}) \ge 0. \label{apeq:cent-pgbt}
\end{gather}
\end{subequations}

If $(p_{g,i}^{*}, \forall i; p_{b,i}^{*}, \forall i; p_{b,g}^{*}, \forall b; p_{g,i}^{'},\forall i; p_{b,i}^{'}, \forall i;  p_{b,g}^{'},\forall b)$ is a SE associated with price ($\lambda_{g,i},\forall i; \lambda_{b,i}, \forall i; \lambda_{b},\forall b$), then according to \eqref{eq:optimality1-2} and \eqref{eq:optimality1-3}, we have $\lambda_{g,i,t}$ must equal $\lambda_{b,i,t}$ for all $i \in \mathcal{I}$ and $t \in \mathcal{T}$. According to \eqref{eq:optimality3-3}, we have $\lambda_{b,t}=\phi_{b,t}^*$ for all $b \in \mathcal{B}$ and $t \in \mathcal{T}$.

To be specific, take \eqref{eq:optimality1-2} as an example, it says that
$\forall\left\{p_{g,i,t},\forall t\right\}\in\mathbb{R}^T,\ (-\lambda_{g,i,t}\ +\eta_{i,t}^\ast)(p_{g,i,t}-p_{g,i,t}^\ast)\geq0,\forall t.$
If we pick up a $p_{g,i,t}>p_{g,i,t}^\ast$, then we have $\eta_{i,t}^\ast\geq\lambda_{g,i,t}$. If we pick up a $p_{g,i,t}<p_{g,i,t}^\ast$, then we have $\eta_{i,t}^\ast\le\lambda_{g,i,t}$. Therefore, $\lambda_{g,i,t}=\eta_{i,t}^\ast$. Similarly, based on \eqref{eq:optimality1-3} we have $\lambda_{b,i,t}=\eta_{i,t}^\ast$. Therefore, $\lambda_{g,i,t}=\eta_{i,t}^\ast=\ \lambda_{b,i,t},\forall i,\forall t$.
\eqref{eq:optimality3-3} says that $\forall\left\{p_{b,g,t},\forall b,\forall t\right\}\in\mathbb{R}^{B\times T},\ (-\lambda_{b,t}\ +\phi_{b,t}^\ast)(p_{b,g,t}-p_{b,g,t}^\ast)\geq0,\forall b,\forall t.$
If we pick up a $p_{b,g,t}>p_{b,g,t}^\ast$, then we have $\phi_{b,t}^\ast\geq\lambda_{b,t}$; If we pick up a  $p_{b,g,t}<p_{b,g,t}^\ast$, then we have $\phi_{b,t}^\ast\le\lambda_{b,t}$. Therefore, $\phi_{b,t}^\ast=\lambda_{b,t}$.

If we let
\begin{subequations}
\begin{align}
    \bar p_{g,i,t}=p_{g,i,t}^*=~ & p_{g,i,t}^{'} ,\forall i,\forall t, \\
    \bar p_{b,i,t}=p_{b,i,t}^*=~ & p_{b,i,t}^{'} ,\forall i,\forall t,\\
    \bar p_{b,g,t}=p_{b,g,t}^{*}=~ & p_{b,g,t}^{'},\forall b,\forall t,\\
    \bar \lambda_{i,t} = \eta_{i,t}^*= \lambda_{g,i,t}=~ &\lambda_{b,i,t},\forall i,\forall t,\\
   \bar\mu_{b,t}= \phi_{b,t}^{*}=~ &  \lambda_{b,t},\forall b,\forall t,
\end{align}
\end{subequations}
then \eqref{apeq:cso-pdit} becomes \eqref{apeq:cent-pdit}, \eqref{apeq:dso-pgit} becomes \eqref{apeq:cent-pgit}, \eqref{eq:optimality2} becomes \eqref{apeq:cent-pbit}, \eqref{apeq:cso-bala} becomes \eqref{apeq:cent-bala}, and  \eqref{apeq:seso-pgbt} becomes \eqref{apeq:cent-pgbt}. Therefore, the system equilibrium is the optimal solution of \eqref{eq:central} with the prices equal to its dual variables.

Similarly, if $(\bar p_{g,i}, \forall i; \bar p_{b,i}, \forall i; \bar p_{b,g}, \forall b)$ is the optimal solution of \eqref{eq:central} and $(\bar \lambda_{i},\forall i; \bar \mu_{b}, \forall b)$ the corresponding dual variable. If we let
\begin{subequations}
\begin{align}
    p_{g,i,t}^*=~ & p_{g,i,t}^{'} = \bar p_{g,i,t},\forall i,\forall t, \\
    p_{b,i,t}^*=~ & p_{b,i,t}^{'} = \bar p_{b,i,t},\forall i,\forall t,\\
    p_{b,g,t}^{*}=~ & p_{b,g,t}^{'} =\bar p_{b,g,t},\forall b,\forall t,\\
    \lambda_{g,i,t}=\lambda_{b,i,t}= ~ & \eta_{i,t}= \bar \lambda_{i,t},\forall i,\forall t,\\
    \lambda_{b,t}=~ &\phi_{b,t}^{*}= \bar\mu_{b,t} ,\forall b,\forall t,
\end{align}
\end{subequations}
then, the optimality conditions \eqref{eq:optimality1}, \eqref{eq:optimality2}, \eqref{eq:optimality3} and the supply-demand balance \eqref{eq:supply-demand} are all satisfied, so a SE is constructed.
 \hfill$\blacksquare$

\setcounter{definition}{0} 
\renewcommand{\theequation}{B.\arabic{equation}}
\renewcommand{\thetheorem}{B.\arabic{theorem}}
\renewcommand{\thedefinition}{B\arabic{definition}}

\section{Proof of Theorem \ref{thm-1}}
\label{appendix-B}
\textit{Proof:}
Let
\begin{align}
    x=~ & \{p_{d,i,t},\forall i\in \mathcal{I},\forall t \in \mathcal{T}; \textbf{0}_{BT \times 1}\} \nonumber\\
    y=~ & \{p_{b,i,t},\forall b \in \mathcal{B},\forall i \in \mathcal{I}_b,\forall t \in \mathcal{T}; p_{b,g,t},\forall b \in \mathcal{B}, \forall t \in \mathcal{T}\} \nonumber\\
    z=~ & \{p_{g,i,t},\forall i \in \mathcal{I},\forall t \in \mathcal{T}; p_{g,b,t},\forall b\in \mathcal{B},\forall t \in \mathcal{T}\} \nonumber\\
    \xi=~ & \{\lambda_{i,t},\forall i \in \mathcal{I},\forall t \in \mathcal{T}; \mu_{b,t},\forall b\in \mathcal{B},\forall t \in \mathcal{T}\} \nonumber\\
    p_{pv}=~ & \{p_{pv,i,t},\forall i \in \mathcal{I},\forall t \in \mathcal{T}; \textbf{0}_{BT \times 1}\}
\end{align}
and
\begin{align}
    \theta_1(x)=~ & \sum_{i \in \mathcal{I}} C_{cs,i} \nonumber\\
    \theta_2(y)=~ & \sum_{b \in \mathcal{B}} C_{bat,b} \nonumber\\
    \theta_3(z)= ~ & C_{ds}
\end{align}
and
\begin{align}
    \mathcal{X}:=\cup_i \mathcal{D}_i,~ \mathcal{Y}:=\mathcal{P}, \mathcal{Z}:=\cup_b \mathcal{F}_b
\end{align}

Let $w=(x, y, z,\xi)$, $\kappa=(y,z,\xi)$ and $u=(x,y,z)$. Then the update \eqref{equ:pditbar}-\eqref{equ:mubar} can be rewritten as
\begin{align}
   \forall x \in \mathcal{X}:& ~ \theta_1(x)-\theta_1(\tilde x^k)+(x-\tilde x^k)^{\top} (-\tilde \xi^k) \ge 0 \\
    \forall y \in \mathcal{Y}:& ~\theta_2(y)-\theta_2(\tilde y^k)   \nonumber\\
    ~ & +(y-\tilde y^k)^{\top} [-\tilde \xi^k + \beta (\tilde y^k-y^k)] \ge 0 \\
    \forall z \in \mathcal{Z}:& ~\theta_3(z)-\theta_3(\tilde z^k) + (z-\tilde z_k)^{\top} \nonumber\\
    ~ & [-\tilde \xi^k +\beta (\tilde y^k-y^k)+\beta(\tilde z^k-z^k)] \ge 0 \\
    \forall \xi \in \mathbb{R}^{I \times T}:&~ (\xi-\tilde \xi^k)^{\top} [\tilde x^k+ y^k+z^k-p_{pv}  \nonumber\\
    ~ & +\frac{1}{\beta}(\tilde \xi^k-\xi^k)] \ge 0
\end{align}
which is equivalent to
\begin{align} \label{eq:variational-update1}
    ~ & \tilde w^k \in \Omega, ~\theta(u)-\theta(\tilde u^{k}) + (w-\tilde w^k)^{\top} F(\tilde w^k) \nonumber\\
    \ge ~ & (\kappa-\tilde \kappa^k)^{\top}Q(\kappa^k-\tilde \kappa^k),~\forall w \in \Omega
\end{align}
where
$\theta(u)=\theta_1(x)+\theta_2(y)+\theta_3(z)$, $\Omega:=\mathcal{X}\times \mathcal{Y} \times \mathcal{Z} \times \mathbb{R}^{(I+B) \times T}$, and
\begin{align}
    F(w)=\left(\begin{array}{c}
         -\xi \\
         -\xi \\
         -\xi \\
         x+y+z-p_{pv} \\
    \end{array}\right)
\end{align}
and
\begin{align}
    Q=\left(\begin{array}{ccc}
        \beta \textbf{I} &  0 & 0 \\
        \beta \textbf{I} &  \beta \textbf{I} &  0 \\
        -\textbf{I} & -\textbf{I} & \frac{1}{\beta} \textbf{I} \\
    \end{array}\right)
\end{align}

Further, the update \eqref{equ:pbitk+1}-\eqref{equ:mubtk+1} can be rewritten as
\begin{align} \label{eq:variational-update2}
    \kappa^{k+1}=\kappa^k-\alpha M (\kappa^k-\tilde \kappa^k)
\end{align}
where
\begin{align}
    M=\left(\begin{array}{ccc}
        \textbf{I} &  -(1-\tau)\textbf{I} & 0  \\
        \tau \textbf{I} & \textbf{I} & 0 \\
        -\beta \textbf{I} & -\beta \textbf{I} & \textbf{I} \\
    \end{array}\right)
\end{align}

Before proving the convergence of the proposed algorithm, we first give the following lemma.

\begin{lemma} \label{lemma-1}
When condition A1 holds, matrices
\begin{align}
    H=QM^{-1}, ~ G=Q^{\top}+Q-\alpha M^{\top} H M
\end{align}
are positive definite.
\end{lemma}

\emph{Proof of Lemma \ref{lemma-1}}:
First, we analyze the matrix $H=QM^{-1}$. Observing the structure of matrices $Q$ and $M$, we can know that the matrix $H$ has the form of
\begin{align}
    H=\left(\begin{array}{ccc}
        h_1 \textbf{I} & h_2 \textbf{I} & 0 \\
        h_3\textbf{I} &  h_4\textbf{I} & 0 \\
        0 & 0 & h_5\\
    \end{array}\right) 
\end{align}
Therefore, we have
\begin{align}
    h_1+\tau h_2 = ~ & \beta \\
    -(1-\tau)h_1+h_2 =~ &0\\
    h_3+\tau h_4 =~ & \beta\\
    -(1-\tau) h_3+h_4 =~ & \beta \\
    h_5 =~ & 1/\beta
\end{align}
Hence,
$h_1=\frac{\beta}{1+\tau(1-\tau)},~ h_2=\frac{(1-\tau)\beta}{1+\tau(1-\tau)}, h_3=\frac{(1-\tau)\beta}{1+\tau(1-\tau)},h_4=\frac{(2-\tau)\beta}{1+\tau(1-\tau)},   h_5=\frac{1}{\beta}$. Since $h_5 \ge 0$, to prove $H$ is a positive definite matrix is equivalent to proving that the following matrix is positive definite:
\begin{align}
    \left(\begin{array}{cc}
        1 &  1-\tau\\
        1-\tau & 2-\tau
    \end{array}\right)\frac{\beta}{1+\tau(1-\tau)}
\end{align}
Since $\tau \in [0,1]$, $\beta>0$, and
\begin{align}
    2-\tau-(1-\tau)^2 = 1+\tau(1-\tau)>0
\end{align}
We have proved that $H$ is a positive definite matrix.

Then, let's consider matrix $G$, which equals to
\begin{small}
\begin{align}
    G = ~ & Q^{\top}+Q-\alpha M^{\top} Q M^{-1} M \nonumber\\
    = ~ & \beta \left(\begin{array}{ccc}
        \textbf{I} & 0 & 0  \\
        0 & \textbf{I} & 0\\
        0 & 0 & \frac{1}{\beta} \textbf{I}
    \end{array}\right) \nonumber\\
    ~ & \left(\begin{array}{ccc}
        2-2\alpha-\alpha\tau  & 1-\alpha-\alpha \tau  & -1+\alpha \\
        1-\alpha-\alpha \tau  & 2-2\alpha & -1+\alpha \\
        -1+\alpha & -1+\alpha & {2-\alpha} \\
    \end{array}\right)  \left(\begin{array}{ccc}
        \textbf{I} &  0 & 0 \\
        0 &  \textbf{I} & 0 \\
        0 & 0 & \frac{1}{\beta}\textbf{I}
    \end{array}\right)
\end{align}
\end{small}
Therefore, $G$ is positive definite if and only if
\begin{align}
    \left(\begin{array}{ccc}
        2-2\alpha-\alpha\tau  & 1-\alpha-\alpha \tau  & -1+\alpha \\
        1-\alpha-\alpha \tau  & 2-2\alpha & -1+\alpha \\
        -1+\alpha & -1+\alpha & {2-\alpha} \\
    \end{array}\right) 
\end{align}
is positive definite, which is condition A1.
\hfill$\blacksquare$

The updates \eqref{eq:variational-update1}, \eqref{eq:variational-update2} have the same form as the prototype algorithm in \cite{he2018class} and according to Lemma \ref{lemma-1}, we can immediately know that the algorithm will converge to the optimal solution of the following optimization problem:
\begin{align}
    \min_{x,y,z}~ & \theta_1(x)+\theta_2(y)+\theta_3(z) \nonumber\\
    \mbox{s.t.}~ & x+y+z=p_{pv} \nonumber\\
    ~ & x \in \mathcal{X}, y\in \mathcal{Y}, z \in \mathbb{Z}
\end{align}
which is the problem \eqref{eq:central}. This completes our proof.  \hfill$\blacksquare$

\bibliographystyle{IEEEtran}
\bibliography{PaperRef}

\end{document}